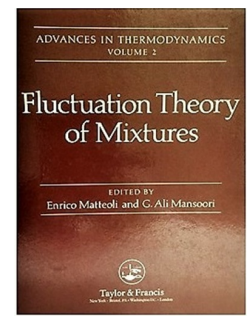

# Relations between and Estimations of Fluctuation Integrals and Direct Correlation Function Integrals

**Luciano Lepori** and **Enrico Matteoli**
*Istituto di Chimica Quantistica ed Energetica Molecolare del CNR, Pisa, Italy*
and
**Esam Z. Hamad** and **G.Ali Mansoori**
*University of Illinois at Chicago (MIC 063), Chicago, IL 60607-7052 USA*

## ABSTRACT

New relations among the mixture direct correlation function integrals (or fluctuation integrals) in terms of concentration variables are developed. These relations indicate that, for example, for a binary mixture only one of the three direct correlation function integrals (or one of the three fluctuation integrals) is independent. Different closure expressions for mixture cross direct correlation function integrals are suggested and they are joined with the exact relations to calculate all the direct correlation function integrals and fluctuation integrals in a mixture.

The results indicate the possibility of introduction of simple closure expressions relating unlike- and like-interaction direct correlation integrals. It is demonstrated that the relation between the direct correlation integrals of hard-sphere mixtures can be satisfactorily represented by a simple geometric mean closure. For Lennard-Jones mixtures with the same intermolecular energy parameters ($\varepsilon_{12}/\varepsilon_{22}=1$) simple geometric mean closure is sufficient for the relation between the direct correlation function integrals so far as values of $\sigma_{12}/\sigma_{22}<1$, while the weighted arithmetic mean closure is necessary for $\sigma_{12}/\sigma_{22}>1$. Generally, a weighted arithmetic mean closure is sufficient in order to represent the relation between the direct correlation function integrals of real (simple or complex) mixtures.

## INTRODUCTION

The knowledge about fluctuation integrals and their relations in mixtures is important in the development of mixture theories of complex molecules (Hamad, et al. 1989; Lepori and Matteoli 1988; Matteoli and Lepori 1984). Studies on concentration fluctuations in mixtures have resulted in numerous advances in the

**Email addresses of authors**: LL (*luciano@indigo.icqem.pi.cnr.it*); EM (*matteoli@ipcf.cnr.it*); EZH (*esam.hamad@aramco.com*); GAM (*mansoori@uic.edu*)

fundamental understanding of behavior of mixtures at, and away from, equilibrium (Gibbs, 1873; Landau and Lifshitz 1980). Definition of properties of mixtures in terms of concentration fluctuations has always been a viable empirical and theoretical route for analysis, correlation and prediction of mixture properties (Hamad, et al. 1987; Kirkwood and Buff 1951; Mansoori and Ely 1985). The engineering concept of "local compositions" is recently linked to the statistical mechanical theory of concentration fluctuations (Mansoori and Ely 1985). Combination of the conformal solution theory and concentration fluctuation theory have resulted in new equation of state mixing rules which have found special utility for analysis and prediction of equilibrium behavior of asymmetric mixtures of interest in such processes as supercritical fluid extraction, retrograde condensation, and miscible flood enhanced oil recovery techniques.

Among the concentration fluctuation theories of statistical mechanics the Kirkwood-Buff solution theory is the one which is developed specifically for mixtures at equilibrium (Kirkwood and Buff 1951). Obtaining thermodynamic properties from this theory requires knowledge of the fluctuation integrals, $G_{ij}$, or the direct correlation function integrals, $C_{ij}$. Evaluating these integrals from microscopic information such as the intermolecular potential energy has not been easy. This is because our knowledge about the intermolecular potential energy functions of real fluids is incomplete and the present techniques for calculating these integrals from potential energy functions are not accurate enough (McQuarrie 1975). Also the relations presented in the first chapter of the present monograph, although exact, can not be used, as they are, to obtain mixture properties. This is because there are too many such integrals, $G_{ij}$'s or $C_{ij}$'s, for a mixture to deal with. In this paper new and exact relations between direct correlation function integrals are reported. Also, since the direct correlation function integrals have simple shapes and short ranges we report here the use of simple combining rules (closures), which are relations between unlike and like interaction direct correlation integrals, for calculation of such integrals and reduction of their number.

## THEORY OF CALCULATING $C_{ij}$'S

In what follows we present the technique for calculating $C_{ij}$'s and $G_{ij}$'s for binary mixtures. Extension of this technique to multicomponent mixtures is straightforward. To derive the relations among the $C_{ij}$ integrals we use the mathematical fact that the mixed second derivatives of a function of two variables are equal at all points where the derivatives are continuous. Let us consider the mole fractions $x_i$, total pressure P, and the absolute temperature T as the independent variables in a mixture in one phase. Then, for the chemical potential, $\mu_i$, of component i in a binary mixture one has (Hamad, et al. 1989):

$$(\partial^2 \mu_i / \partial x_1 \partial P)_T = (\partial^2 \mu_i / \partial P \partial x_1)_T \qquad i = 1,2 \qquad (1)$$

The derivative on the right-hand side of the above equation is equal to $(\partial \underline{v_i} / \partial x_1)_{T,P}$

becomes

$$(\partial^2 \mu_i / \partial x_1 \partial P)_T = (\partial \bar{V}_i / \partial x_1)_{T,P} \qquad i = 1,2 \qquad (2)$$

Substituting the expressions for the partial molar volume and the chemical potential with respect to the direct correlation function integrals $C_{ij}$ (see Chapter 1),

$$\rho \bar{V}_i = \left(1 - \rho \sum_{j=1}^{2} x_j C_{ij}\right) \bigg/ \left(1 - \rho \sum_{j=1}^{2} \sum_{k=1}^{2} x_j x_k C_{jk}\right) \qquad (3)$$

$$x_1 [\partial \mu_1 / \partial x_1]_{P,T} = kT \frac{1 - x_1 \rho C_{11} - x_2 \rho C_{22} + x_1 x_2 \rho^2 (C_{11} C_{22} - C_{12}^2)}{1 - \Sigma \Sigma x_i x_j \rho C_{ij}} \qquad (4)$$

in Eq. (2) gives the following two equations:

$$kT \partial \{[1 - x_1 \rho C_{11} - x_2 \rho C_{22} + x_1 x_2 \rho^2 (C_{11} C_{22} - C_{12}^2)] / [1 - \Sigma \Sigma x_i x_j \rho C_{ij}]\} / \partial p$$
$$= x_1 \partial \{[1 - x_1 \rho C_{11} - x_2 \rho C_{12}] / [\rho (1 - \Sigma \Sigma x_i x_j \rho C_{ij})]\} / \partial x_1 \qquad (5)$$

and

$$kT \partial \{[1 - x_1 \rho C_{11} - x_2 \rho C_{22} + x_1 x_2 \rho^2 (C_{11} C_{22} - C_{12}^2)] / [1 - \Sigma \Sigma x_i x_j \rho C_{ij}]\} / \partial p$$
$$= x_2 \partial \{[1 - x_2 \rho C_{22} - x_1 \rho C_{12}] / [\rho (1 - \Sigma \Sigma x_i x_j \rho C_{ij})]\} / \partial x_2 \qquad (6)$$

These two equations constitute two independent expressions relating $C_{11}$, $C_{22}$ and $C_{12}$ in a binary mixture. In principle Eq.s (5) and (6), together with another expression relating $C_{12}$ to $C_{11}$ and $C_{22}$, can be used to solve for the three quantities $C_{11}$, $C_{22}$ and $C_{12}$.

<u>An Alternative Derivation:</u> Consider a system which is described by the following set of independent variables $\{T, \rho_1, \rho_2, \ldots, \rho_n\}$, where T is the absolute temperature, $\rho_i = N_i / V$, $N_i$ is the number of molecules i in the system, V is the system volume, and n is the number of components in the mixture. In the grand canonical ensemble the following expression for $\partial \mu_i / \partial \rho_j$ is derived (O'Connell 1971; Landau and Lifshitz 1980)

$$[\partial \mu_i / \partial \rho_j]_{T, \rho_j'} = kT \{\delta_{ij} / \rho_i - C_{ij}\} \qquad (7)$$

where $\rho_j'$ stands for the set $\rho_1, \rho_2, \ldots, \rho_n$ variables excluding $\rho_j$,

$$C_{ij} = \int c_{ij}(r) dr \qquad (8)$$

and $c_{ij}(r)$ is the direct correlation function of species i and j. Using Eq. (7) exact relations can be derived among the direct correlation function integrals, $C_{ij}$, by differentiating this equation with respect to $\rho_i$ and equating the mixed second derivatives of the chemical potentials (Hamad, et al. 1989). For example, for a binary mixture with T, $\rho_1, \rho_2$ as the independent variables one can write:

$$\partial C_{12}/\partial \rho_1 = \partial C_{11}/\partial \rho_2 \tag{9}$$

and

$$\partial C_{12}/\partial \rho_2 = \partial C_{22}/\partial \rho_1 \tag{10}$$

The above two equations are exact relating $C_{11}$, $C_{22}$, and $C_{12}$. In order to solve these equations for $C_{11}$, $C_{22}$, and $C_{12}$ one needs a closure relation among $C_{11}$, $C_{22}$, and $C_{12}$ to solve for the three direct correlation function integrals. In the next section a number of closure expressions for $C_{12}$ are considered and the direct correlation integrals are solved for the hard-sphere fluid mixture, the Lennard-Jones fluid mixture, and real mixtures consisting of simple and complex molecules.

## CHOICE OF THE $C_{12}$ CLOSURE

Simultaneous solution of Eq.s (5) and (6) for $C_{11}$, $C_{22}$ and $C_{12}$ is not possible without another expression relating $C_{12}$ to $C_{11}$ and $C_{22}$ so that

$$C_{12} = C_{12}(C_{11}, C_{22}; T, p, x_1, x_2, \ldots x_{n-1}) \tag{11-1}$$

Similarly, simultaneous solution of Eq.s (9) and (10) for $C_{11}$, $C_{22}$ and $C_{12}$ is not possible without another expression relating $C_{12}$ to $C_{11}$ and $C_{22}$ so that

$$C_{12} = C_{12}(C_{11}, C_{22}; T, \rho_1, \rho_2, \ldots, \rho_n) \tag{11-2}$$

We choose to call Eq (11-1), or (11-2), the $C_{12}$ closure. The exact $C_{12}$ closure expression is not presently available. There exist variety of ways of assuming the closure for the direct correlation function integrals. With an appropriate choice for the functional form of the closure expression, Eq.s (5, 6, 11-1), or Eq.s (9, 10, 11-2), can be solved simultaneously for $C_{11}$, $C_{22}$ and $C_{12}$. Then the fluctuation integrals $G_{11}$, $G_{22}$ and $G_{12}$ can be derived from $C_{11}$, $C_{22}$ and $C_{12}$ using the following equations (Pearson and Rushbrooke 1957; O'Connell 1971, 1981).

$$\rho G_{11} = \frac{\rho C_{11} - (1-x_1)\rho^2(C_{11}C_{22} - C_{12}^2)}{1 - x_1\rho C_{11} - x_2\rho C_{22} + x_1 x_2 \rho^2(C_{11}C_{22} - C_{12}^2)} \tag{12}$$

$$\rho G_{22} = \frac{\rho C_{22} - (1 - x_2)\rho^2(C_{11}C_{22} - C_{12}^2)}{1 - x_1\rho C_{11} - x_2\rho C_{22} + x_1 x_2 \rho^2(C_{11}C_{22} - C_{12}^2)} \quad (13)$$

$$\rho G_{12} = \frac{\rho C_{12}}{1 - x_1\rho C_{11} - x_2\rho C_{22} + x_1 x_2 \rho^2(C_{11}C_{22} - C_{12}^2)} \quad (14)$$

Before making any assumption about the mathematical expression of the $C_{ij}$ closure it is helpful to study the behavior of certain known direct correlation function integrals.

The direct correlation functions for a mixture of hard-spheres have been obtained from the Percus-Yevick approximation (Lebowitz 1964). The direct correlation function integrals were then calculated from Lebowitz's work (Hamad 1988). It was found that for a binary mixture of hard-spheres $C_{12}$ always lies in between $C_{11}$ and $C_{22}$. This suggests that $C_{12}$ can be approximated by some kind of an average of $C_{11}$ and $C_{22}$

Another source of information about the mathematical form of the $C_{ij}$ closure is based on the fact that the virial expansion of the direct correlation function integral can be written in the following form (Hamad 1988)

$$C_{ij} = -\left(2B^{(2)}_{ij} + 3\sum_{k=1}^{n} B^{(3)}_{ijk}\rho_k + 4\sum_{k=1}^{n}\sum_{\ell=1}^{n} B^{(4)}_{ijk\ell}\rho_k\rho_\ell + \dots \right) \quad (15)$$

where $B^{(2)}_{ij}$, $B^{(3)}_{ijk}$ and $B^{(4)}_{ijk\ell}$ are the second, third and fourth virial coefficients, respectively. From this equation we can conclude,

$$\text{Limit } C_{ij} = -2 B^{(2)}_{ij}(T) \text{ as } \rho \longrightarrow 0, \quad (16)$$

where $B^{(2)}_{ij}(T)$ is the second virial coefficient of the pair i and j. At low densities the sign of $C_{ij}$ is opposite of that of the second virial coefficient, $B^{(2)}_{ij}$. It is well known that, while the second virial coefficient of hard-sphere fluids is always positive, the second virial coefficient of real fluids can be positive or negative depending on the temperature. This implies that $C_{11}$, $C_{22}$ and $C_{12}$ of real mixtures can have positive or negative signs.

Considering the above facts we can assume $C_{12}$ is an average of $C_{11}$, $C_{22}$. Then, some of the possible expressions which may be chosen for Eq. (11) are as the following:

**1) Arithmetic mean closure:** By postulating that $C_{12}$ is an arithmetic mean of $C_{11}$ and $C_{22}$ one may write

$$C_{12} = (C_{11} + C_{22})/2 \tag{17}$$

Eq. (17) implies that the mixture under consideration is an ideal solution (Hamad 1988). Instead of Eq. (17) one may write the following weighted arithmetic mean $C_{12}$ closure for non-ideal solutions (Hamad, et al. 1989)

$$C_{12} = \alpha_{21}C_{11} + \alpha_{12}C_{22} \tag{18}$$

Parameters $\alpha_{21}$ and $\alpha_{12}$ are adjustable interaction parameters which have to be determined for the interacting species under consideration. By joining Eq.s (16) and (18) and choosing $\alpha_{21}\alpha_{12}=1/4$ (for the sake of simplicity) we can derive the following limiting expression for $\alpha_{21}$ with respect to the second virial coefficients.

$$\alpha_{21} = B_{12}/B_{11} \pm \left[(B_{12}/B_{11})^2 - B_{22}/B_{11}\right]^{1/2} \tag{16-1}$$

As it is demonstrated by Hamad and Mansoori (1989) the choice of $\alpha_{21}\alpha_{12}=1/4$ allows us to derive analytic expressions for chemical potentials of components in the mixture.

2) <u>Geometric mean closure:</u> This closure is suggested based on a relation among the contact values of the direct correlation functions in a mixture of one-dimensional hard-rods. It is shown that the contact $c_{ij}$ values are related by (Lebowitz 1964)

$$c_{12}^2(\sigma_{12}) = c_{11}(\sigma_{11}) \, c_{22}(\sigma_{22}) \tag{19}$$

where $\sigma_{ij}$ is the distance of closest approach between centers of particles i and j, and $\sigma_{ij}=(\sigma_{ii}+\sigma_{jj})/2$. For the direct correlation integrals of a mixture of hard-spheres a relation similar to Eq. (19) may be assumed (Hamad, et al. 1989)

$$C_{12}^2 = C_{11}C_{22} \tag{20}$$

In this equation $C_{12}$ is assumed to be the geometric mean of $C_{11}$ and $C_{22}$. For molecules other than hard-spheres Eq. (20) may be replaced with the following expression:

$$C_{12}^2 = \beta \, C_{11}C_{22}, \tag{21}$$

where $\beta$ is an adjustable interaction parameter.

3) <u>A combined arithmetic and geometric mean closure:</u> In the above mentioned closures parameters $\alpha_{21}$, $\alpha_{12}$, and $\beta$ are adjustable parameters. In order to make the closure more versatile (with more adjustable parameters) one may combine the arithmetic and geometric means in order to produce the following closure

$$C_{12} = \alpha_{21}C_{11} + \alpha_{12}C_{22} \pm (\beta C_{11}C_{22})^{1/2}, \qquad (22)$$

However, as it will be demonstrated later for all the binary mixtures studied here, either of the geometric or arithmetic mean closures are sufficient enough to represent $C_{12}$ with respect to $C_{11}$ and $C_{22}$. It is also possible to come up with variety of other $C_{12}$ closure expressions. However, the above mentioned $C_{12}$ closures are the simplest possible relations that one may suggest. The simplicity of these closures makes it easy to use them in calculating the $G_{ij}$ and $C_{ij}$ integrals.

## TEST OF $C_{12}$ CLOSURES

1) <u>hard-Sphere Mixture:</u> The validity of the proposed closures can be tested by comparing the fluctuation integrals or direct correlation function integrals calculated by the present technique with their exact values. For this purpose we first consider a model fluid mixture (hard-sphere mixture) for which accurate direct correlation function integrals are already available. The Percus-Yevick theory (Percus and Yevick 1958) gives quite accurate expressions for the hard-sphere mixture direct correlation function integrals (Lebowitz 1964). In Table 1 the $C_{12}/C_{11}$ values calculated by the Percus-Yevick theory are compared with the result of geometric mean closure, Eq. (20) and the arithmetic mean closure, Eq. (17) for a binary equimolar mixture of hard-spheres with $\sigma_{11}/\sigma_{22}=1.5$.

Table 1. Values of $C_{12}/C_{11}$ for an equimolar hard-sphere mixture ($\sigma_{11}/\sigma_{22}=1.5$) calculated according to simple geometric mean (SGM), Eq. (20), and weighted arithmetic mean (WAM), Eq. (18), closures and compared with the Percus-Yevick (P-Y) theory (Hamad and Mansoori 1989).

| $(\pi/6)\rho\sigma_{22}^3$ | P-Y | SGM | % Dev. | WAM | % Dev. |
|---|---|---|---|---|---|
| 0.05 | 0.5265 | 0.5047 | 4.1 | 0.5243 | 0.4 |
| 0.10 | 0.4806 | 0.4677 | 2.7 | 0.4772 | 0.7 |
| 0.20 | 0.4066 | 0.4034 | 0.8 | 0.4039 | 0.7 |
| 0.30 | 0.3530 | 0.3526 | 0.1 | 0.3537 | -0.2 |
| 0.40 | 0.3141 | 0.3142 | 0.02 | 0.3202 | -1.9 |
| 0.50 | 0.2844 | 0.2844 | 0.0 | 0.2969 | -4.4 |

For the weighted arithmetic mean closure, $\alpha_{21}\alpha_{12}=1/4$ is chosen. At $\sigma_{11}/\sigma_{22}=1.5$, the negative sign in Eq. (16-1) gives $\alpha_{21}=0.1911$. See Hamad and Mansoori (1989).

---

This table shows clearly that both geometric and arithmetic mean closures are accurate enough for hard-sphere mixture. However, the geometric mean closure is more accurate at high densities while the weighted arithmetic mean closure is more accurate at low densities.

2) <u>Real Mixtures</u>: Data on fluctuation integrals are available for variety of binary mixtures (Matteoli and Lepori, 1984; Lepori and Matteoli, 1988). Such data can be introduced into inverted Eqs. (12)-(14) to calculate $C_{ij}$ and then used for the test of different closure expressions. The validity of $C_{12}$ closures can be tested for real mixtures by plotting $C_{12}/C_{11}$ data versus $C_{22}/C_{11}$ data and studying the slope of the resulting curve. Figures 1 and 2 show the plot of $C_{12}/C_{11}$ versus $C_{22}/C_{11}$ for four different binary mixtures. The linearity of these plots suggests that the weighted arithmetic mean closure, Eq. (18), is quite sufficient for real mixtures. As a further

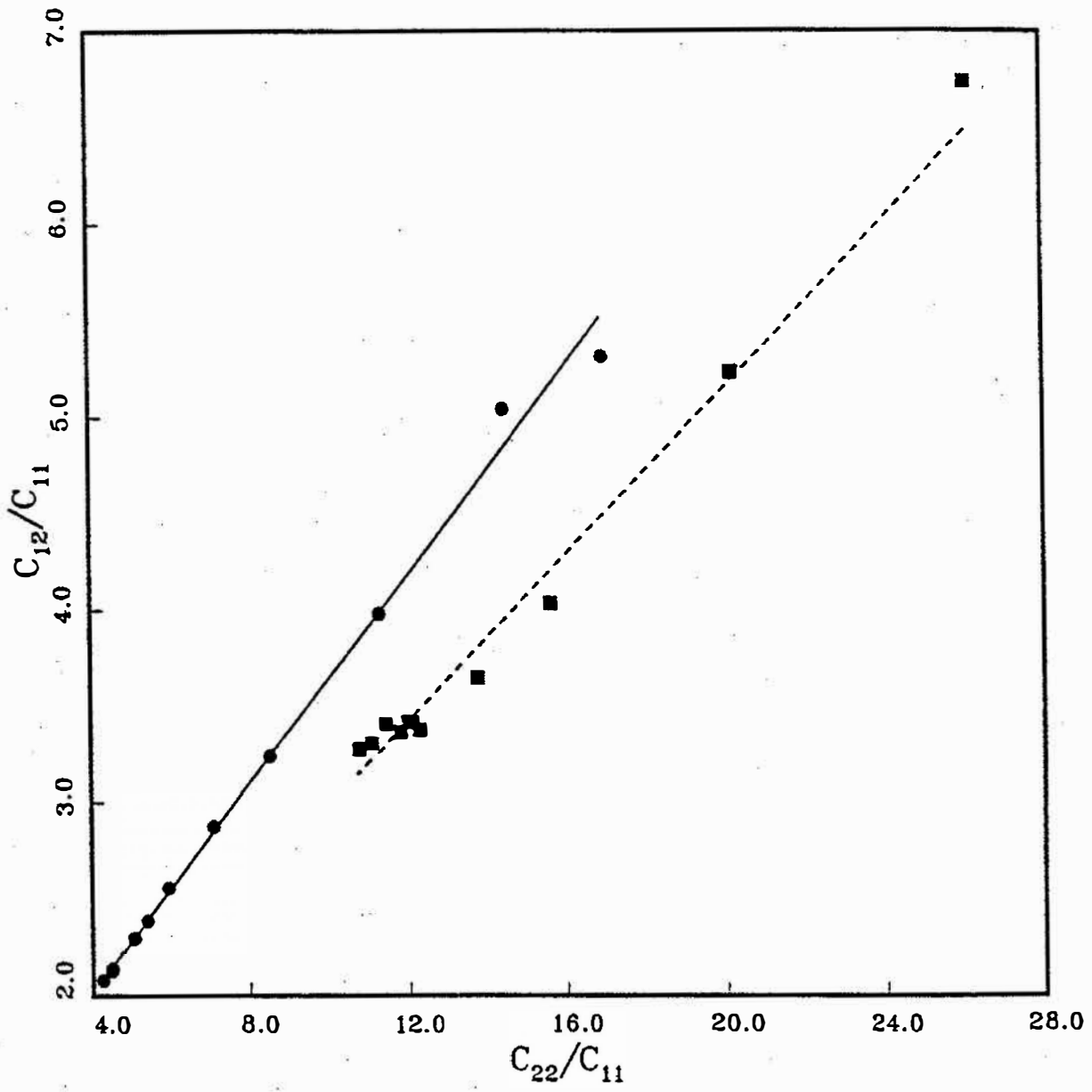

Figure 1. The test of the linearity of $C_{12}/C_{11}$ versus $C_{22}/C_{11}$ for water(1) + methanol(2) (circles) and water (1) + aminoethanol (2) (squares) binary mixtures. The $C_{ij}$ data has been calculated from $G_{ij}$ values taken from Matteoli and Lepori (1984). The curves represent the weighted arithmetic mean closure, Eq. (18).

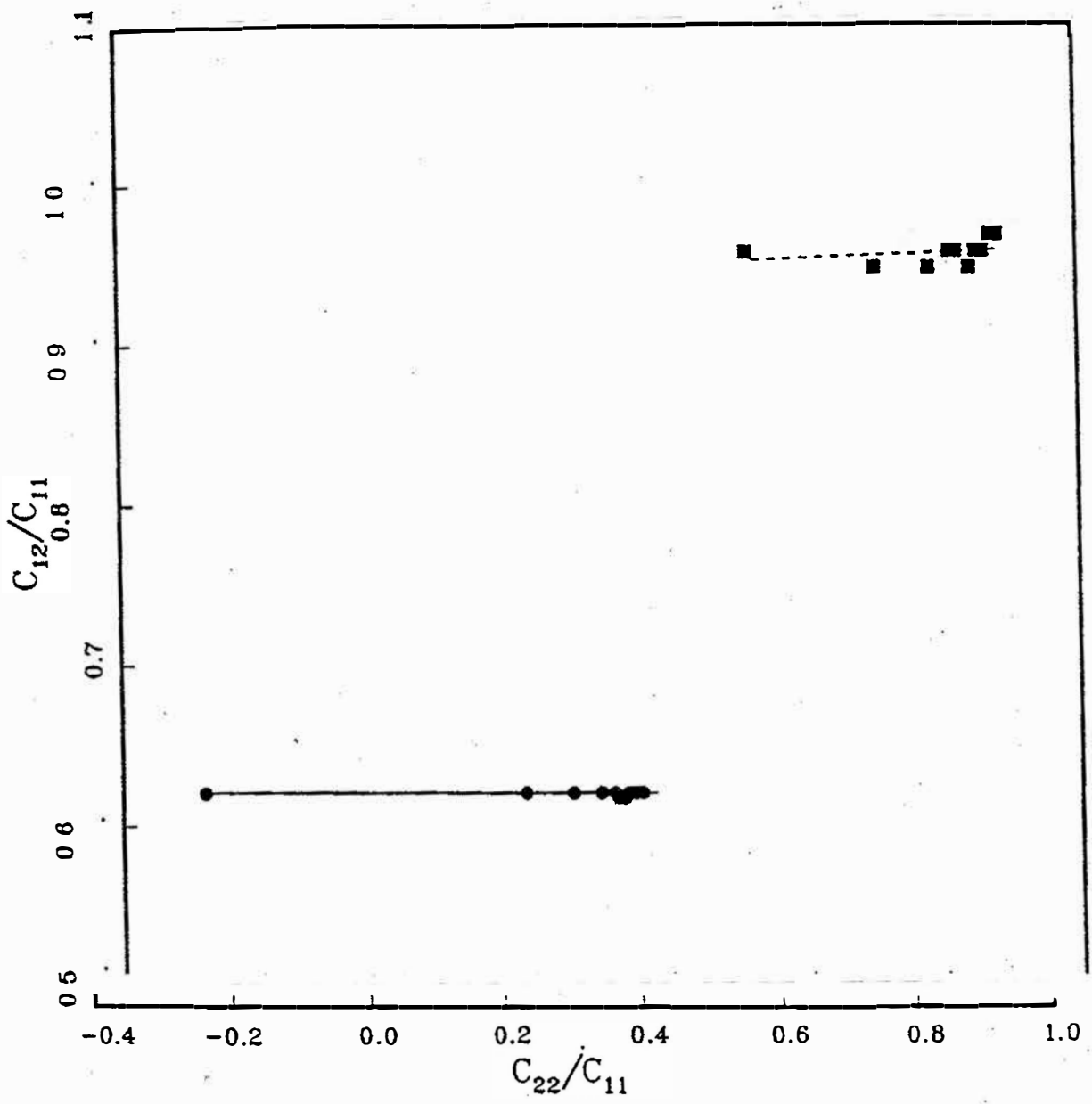

Figure 2. The test of the linearity of $C_{12}/C_{11}$ versus $C_{22}/C_{11}$ for tetrachloromethane (1) + ethanol (2) (circles) and tetrachloromethane (1) + n-butanol (2) (squares) binary mixtures. The $C_{ij}$ data has been calculated from $G_{ij}$ values taken from Lepori and Matteoli (1988). The curves represent the weighted arithmetic mean closure, Eq. (18).

Table 2. The deviations of real fluid direct correlation integrals from the relation
$C_{12} = \alpha_{21} C_{11} + \alpha_{12} C_{22}$

| System | Temperature [°C] | S X100 | Max Dev. [%] | $x_1$ at Max Dev. |
|---|---|---|---|---|
| water+methanol | 0.0 | 2.1 | 7.7 | 0.0 |
| "         " | 25. | 1.9 | 7.5 | 0.0 |
| "         " | 60. | 2.2 | -2.6 | 0.4 |
| water+ethanol | 25. | 14.5 | 22. | 0.0 |
| "         " | 50. | 3.2 | 4.1 | 0.1 |
| "         " | 90. | 4.1 | 6.7 | 0.1 |
| water+propanol | 25. | 2.7 | 7.4 | 0.0 |
| water+tert-butanol | 25. | 1.5 | -4.6 | 1.0 |
| water+acetonitrile | 30. | 2.3 | -11. | 1.0 |
| water+acetone | 25. | 1.9 | -4.4 | 1.0 |
| water+dimethylsulfoxide | 25. | 5.9 | -8.7 | 0.0 |
| water+tetrahydrofuran | 25. | 4.3 | -8.8 | 1.0 |
| water+1,4-dioxane | 25. | 4.0 | -5.9 | 1.0 |
| water+2-aminoethanol | 25. | 6.9 | -6.2 | 1.0 |
| CCl4+methanol | 25. | 0.32 | -2.4 | 0.0 |
| CCl4+ethanol | 25. | 0.28 | -2.1 | 0.0 |
| CCl4+propanol | 25. | 0.12 | -1.3 | 0.0 |
| CCl4+n-butanol | 25. | 0.12 | -1.2 | 0.0 |
| CCl4+tetrahydrofuran | 25. | 0.0004 | .005 | 0.9 |
| CCl4+1,4-dioxane | 25. | 0.00003 | 0.003 | 0.0 |
| CY-hexane+diethylether | 25. | 0.014 | -0.4 | 1.0 |
| CY-hexane+dipropylether | 25 | 0.030 | -0.6 | 0.0 |
| CY-hexane+dibuthylether | 25. | 0.002 | 0.2 | 0.0 |
| CY-hexane+ethylbutylether | 25. | 0.020 | -0.5 | 0.0 |
| CY-hexane+dimethoxymethane | 25. | 0.24 | -2.1 | 1.0 |
| CY-hexane+diethoxymethane | 25. | 0.0005 | -0.02 | 1.0 |
| CY-hexane+diethoxyethane | 25. | 0.19 | -1.8 | 0.0 |
| CY-hexane+diglyme | 25. | 0.16 | 1.0 | 1.0 |

In this table, $S = \Sigma\{(C_{12,pred} - C_{12,exp})/C_{12,exp}\}^2$ and $\%Dev. = \{(C_{12,pred} - C_{12,exp})/C_{12,exp}\} \times 100$.

Experimental $C_{ij}$ values were evaluated from experimental $G_{ij}$'s using inverted Eq's (12)-(14). $G_{ij}$'s were taken from Matteoli and Lepori (1984), Lepori and Matteoli (1988), or calculated using activity coefficient data (Lepori, et al. 1988). Density data were taken from Berti, et al., (1988) and isothermal compressibility data were taken from Brostow and Maynadier (1979) and Handbook of Chemistry and Physics (1989).

The maximum deviation for all studied systems is about 22%. This large error happens where the $C_{ij}$ goes to zero and crosses the horizontal axis. Since some of the data has uncertainties up to 30-40% (Lepori and Matteoli, 1988), the present results demonstrate that the weighted arithmetic mean closure is a good closure approximation for real mixtures.

## CACULATION OF MIXTURE $C_{ij}$'S FORM PURE FLUID DATA

To avoid having to solve numerically the nonlinear partial differential equations, Eq.s (5 & 6) or (9 & 10), one can make use of the fact that for liquids at low to moderate pressures the direct correlation function integrals are week functions of pressure. Under this assumption, for example Eq.s (5) and (6) will give:

$$(1-x_1\rho C_{11}-x_2\rho C_{12})\bigg/\bigg(\rho-\rho^2\sum\sum x_j x_k C_{jk}\bigg) = v_1 \tag{23}$$

$$(1-x_1\rho C_{12}-x_2 v C_{22})\bigg/\bigg(\rho-\rho^2\sum\sum x_j x_k C_{jk}\bigg) = v_2 \tag{24}$$

where $v_1$ and $v_2$ are the molar volumes of pure components 1 and 2, respectively. Comparisons of Eq.s (23) and (24) with Eq. (3) imply that the partial molar volumes are set equal to the pure component volumes at all concentrations. This is a result of assuming that $C_{ij}$'s are pressure-independent. This approximation, which is also made in lattice models of liquid mixtures, does not have a pronounced effect on the $C_{ij}$'s as will be seen later. Joining Eq.s (23) and (24) with the equation for the isothermal compressibility,

$$(\rho k T \kappa_T)^{-1} = 1 - \rho \sum_{i=1}^{2}\sum_{j=1}^{2} x_i x_j C_{ij} \tag{25}$$

gives:

$$x_1\rho C_{11} + x_2\rho C_{12} = 1 + W[(v_1/v_2)x_1 + x_2] \tag{26}$$

$$x_1\rho C_{12} + x_2\rho C_{22} = 1 + W[x_1 + (v_2/v_1)x_2] \tag{27}$$

where

$$W = -\left[kT(x_1\kappa_{T_1}/v_2 + x_2\kappa_{T_2}/v_1)\right]^{-1} \tag{28}$$

and $\kappa_{T_1}$ and $\kappa_{T_2}$ are the isothermal compressibilities of the pure components 1 and 2, respectively. The direct correlation integrals, $C_{ij}$'s, in a binary mixture can now be solved for by combining Eq.s (18), (26) and (27) with the following result:

$$\rho C_{11} = \frac{W(1-\alpha_{12}v_2/v_1)x_2^2 + Wx_1(\alpha_{12}x_1+x_2)v_1/v_2 + (1-\alpha_{12})x_2 + \alpha_{12}x_1}{\alpha_{21}x_2^2 + x_1x_2 + \alpha_{12}x_1^2} \tag{29}$$

$$\rho C_{22} = \frac{W(1-\alpha_{21}v_1/v_2)x_1^2 + Wx_2(\alpha_{21}x_2+x_1)v_2/v_1 + (1-\alpha_{21})x_1 + \alpha_{21}x_2}{\alpha_{21}x_2^2 + x_1x_2 + \alpha_{12}x_1^2} \tag{30}$$

$$\rho C_{12} = \frac{W(\alpha_{12}x_1^2 + \alpha_{21}x_1x_2v_1/v_2 + \alpha_{12}x_1x_2v_2/v_1 + \alpha_{21}x_2^2) + \alpha_{12}x_1 + \alpha_{21}x_2}{\alpha_{21}x_2^2 + x_1x_2 + \alpha_{12}x_1^2} \quad (31)$$

With the availability of pure fluid thermodynamic data and interaction parameters $\alpha_{21}$ and $\alpha_{12}$ Eqs (29-31) can be used for calculation of mixture direct correlation function integrals.

## RESULTS AND DISCUSSION

The $C_{12}$ closures discussed in the previous sections can be used to predict the properties of hard-sphere mixtures, Lennard-Jones mixtures and real mixtures using pure fluid properties (Hamad, et al. 1989; Hamad and Mansoori 1989).

For hard-spheres the Carnahan-Starling (1969) equation of state is used to describe the behavior of the pure hard-sphere fluid. In Table 3 the chemical potential predictions using the two simple $C_{12}$ closures, Eq.s (17) and (20), are compared to the Mansoori et al. (1971) hard-sphere mixture equation, which is known to reproduce the computer simulation data accurately.

Table 3. The variation of the chemical potential of component 1 in a mixture of hard-spheres ($\sigma_{11}/\sigma_{22}=1.5$, $\rho\sigma_{22}^3=0.25$) with composition based on different closures and compared with the MCSL (Mansoori, et al. 1971) equation of state (Hamad and Mansoori 1989).

| $x_1$ | | $\mu_{1r}/kT$ | | | |
|---|---|---|---|---|---|
| | MCSL | $C_{12}=(C_{11}C_{22})^{1/2}$ | %Dev. | $C_{12}=\alpha_{21}C_{11}+\alpha_{12}C_{22}$ | %Dev. |
| 0.177 | 3.60 | 3.49 | 2.9 | 3.56 | 1.1 |
| 0.340 | 4.49 | 4.40 | 2.0 | 4.44 | 1.1 |
| 0.491 | 5.53 | 5.46 | 1.3 | 5.47 | 1.1 |
| 0.631 | 6.73 | 6.67 | 0.8 | 6.67 | 0.9 |
| 0.761 | 8.12 | 8.09 | 0.4 | 8.07 | 0.6 |
| 0.883 | 9.75 | 9.74 | 0.2 | 9.72 | 0.3 |

For the weighted arithmetic mean closure $\alpha_{21}\alpha_{12}=1/4$. is chosen. At $\sigma_{11}/\sigma_{22}=1.5$ and the negative sign in Eq. (16-1) gives $\alpha_{21}=0.1911$. See Hamad and Mansoori (1989).

The chemical potential calculations reported in this table are for component 1 of a binary mixture with $\sigma_{11}/\sigma_{22}=1.5$ and $\rho\sigma_{22}^3=0.25$. The dimensionless density $\rho\sigma_{22}^3=0.25$ corresponds to 95% of the solidification density of component 1. Accor-ding to Table 3, for prediction of the hard-sphere mixture chemical potentials either of the two simple $C_{12}$ closures, Eq.s (17) or (20), are quite sufficient. Similar results are also obtained for the chemical potential of component 2, and at other conditions which will not be reported here.

The hard-sphere intermolecular potential energy function is purely repulsive. To test the proposed closure models for mixtures of molecules with intermolecular potentials which possess, both, attractive and repulsive parts, the Lennard-Jones potential model is considered here. The chemical potential of a mixture of Lennard-Jones fluids at infinite dilution is predicted here and it is compared with the available computer simulation data. The necessary pure component Lennard-Jones fluid data is calculated using the Nicolas et al. (1979) Lennard-Jones fluid equation of state. The chemical potential predictions are compared to the infinite dilution simulation data of Shing (1982) in Table 4.

Table 4. The chemical potential at infinite dilution of Lennard-Jones mixture. ($kT/\varepsilon_{22}$ =1.2, $\rho\sigma_{22}^3$=0.7). (Hamad and Mansoori, 1989).

| $(\sigma_{12}/\sigma_{22})^3$ | $\varepsilon_{12}/\varepsilon_{22}$ | $\mu^{\infty}_{1r}/kT$ | | |
|---|---|---|---|---|
| | | Simulation Data | $C_{12}=(C_{11}C_{22})^{1/2}$ | $C_{12}=\alpha_{21}C_{11}+\alpha_{12}C_{22}$ |
| 0.3 | 1.0 | -1.30 | -1.75 | +4.14 |
| 0.5 | 1.0 | -1.63 | -1.80 | -1.29 |
| 0.75 | 1.0 | -1.83 | -1.86 | -1.92 |
| 1.0 | 1.0 | -1.93 | -1.99 | -1.99 |
| 1.5 | 1.0 | -1.87 | -2.26 | -1.65 |
| 2.0 | 1.0 | -1.55 | -2.58 | -1.52 |

In this table, the simulation data are taken from Shing (1982), and Shing et al, (1988).

According to this table for molecules with the same intermolecular energy parameters, $\varepsilon_{12}/\varepsilon_{22}$=1, the weighted arithmetic mean closure shows better agreement with the simulation data at $\sigma_{12}/\sigma_{22}$>1. Geometric mean closure shows better agreement with the simulation data at $\sigma_{12}/\sigma_{22}$<1.

A critical test of the present technique for calculating $C_{ij}$ and $G_{ij}$ integrals is its application to real mixtures. We already have demonstrated that the weighted arithmetic mean closure is sufficient to correlate the direct correlation function integrals of real mixtures (see Table 2). Eqs (5), (6), and (18) are joined together to fit the experimental $C_{ij}$ data by optimizing parameters $\alpha_{21}$ and $\alpha_{12}$. The experimental $C_{ij}$ integrals for real mixtures are calculated from experimental $G_{ij}$ integrals (Matteoli and Lepori, 1984; Lepori and Matteoli, 1988; Lepori, et al., 1988; Berti, et al., 1989; and Hamad, et al., 1989) inverting Eqs (12) - (14).

In Figures 3 - 13 the $C_{11}$, $C_{22}$ and $C_{12}$ integrals resulting from this fit are reported for twenty two different binary mixtures and they are compared with the experimental $C_{11}$, $C_{22}$ and $C_{12}$ data.

According to Figures 3 - 13, the weighted arithmetic mean closure, Eq. (18), is quite sufficient for application to real mixtures. Table 2 lists the results of fitting Equation (18) to twenty-eight binary mixtures that include the twenty-two systems reported in Figures 3-13 and comprise a wide variety of non-polar, polar, and associating compounds. In Table 2 the sum of squares of the deviations, the maximum percentage deviation and the mole fraction of component 1 at the maximum deviation are also reported. According to this table, the maximum deviation for all the systems studied is 22% (water + ethanol mixture at 25 C), and usually it occurs at the infinite dilution region of one of the two components of the binary mixture (far right column in Table 2). Keeping in mind that some of the $C_{ij}$ data have uncertainty of up to 30 - 40% (Lepori and Matteoli 1988) we conclude that Eq. (18) is an excellent closure for the relation between $C_{12}$, $C_{11}$, and $C_{22}$. All the results reported in this table are at constant temperature and at low pressures. It should be pointed out that parameters $\alpha_{21}$ and $\alpha_{12}$ are expected to be dependent on temperature and, to a lower degree, to pressure.

Figures 14-19 represent comparisons of $G_{ij}$ values calculated using the present technique (Eqs. 28-31 and 12-14) and the experimental values for water + organic and tetrachloromethane (TCM) + organic mixtures. The experimental $G_{ij}$ values were taken from Matteoli and Lepori (1984) for the aqueous systems and from Lepori and Matteoli (1988) for mixtures containing TCM. The agreement between the experimental and calculated data ranges from excellent as in the case of TCM+dioxane, to poor as in the case of water+dioxane. The largest deviations are seen at the extrema (maxima or minima) of every curve where the experimental uncertainties are highest at these locations. Overall, according to these figures the calculated $G_{ij}$ integrals are in good agreement with the experimental values.

Table 5 shows the values of the molar volumes and the isothermal compressibilities of the pure components used in Eq.s (29)-(31).

Parameters $\alpha_{21}$ and $\alpha_{12}$ for different binary mixtures are calculated and they are reported in Table 6. These parameters are obtained by minimizing the following objective function

$$\text{O.F.} = \sum |G_{11}/G_{11,ex}-1| + |G_{12}/G_{12,ex}-1| + |G_{22}/G_{22,ex}-1| \qquad (32)$$

where $G_{ij,ex}$ is the experimental $G_{ij}$ and the summation runs over all the experimental points.

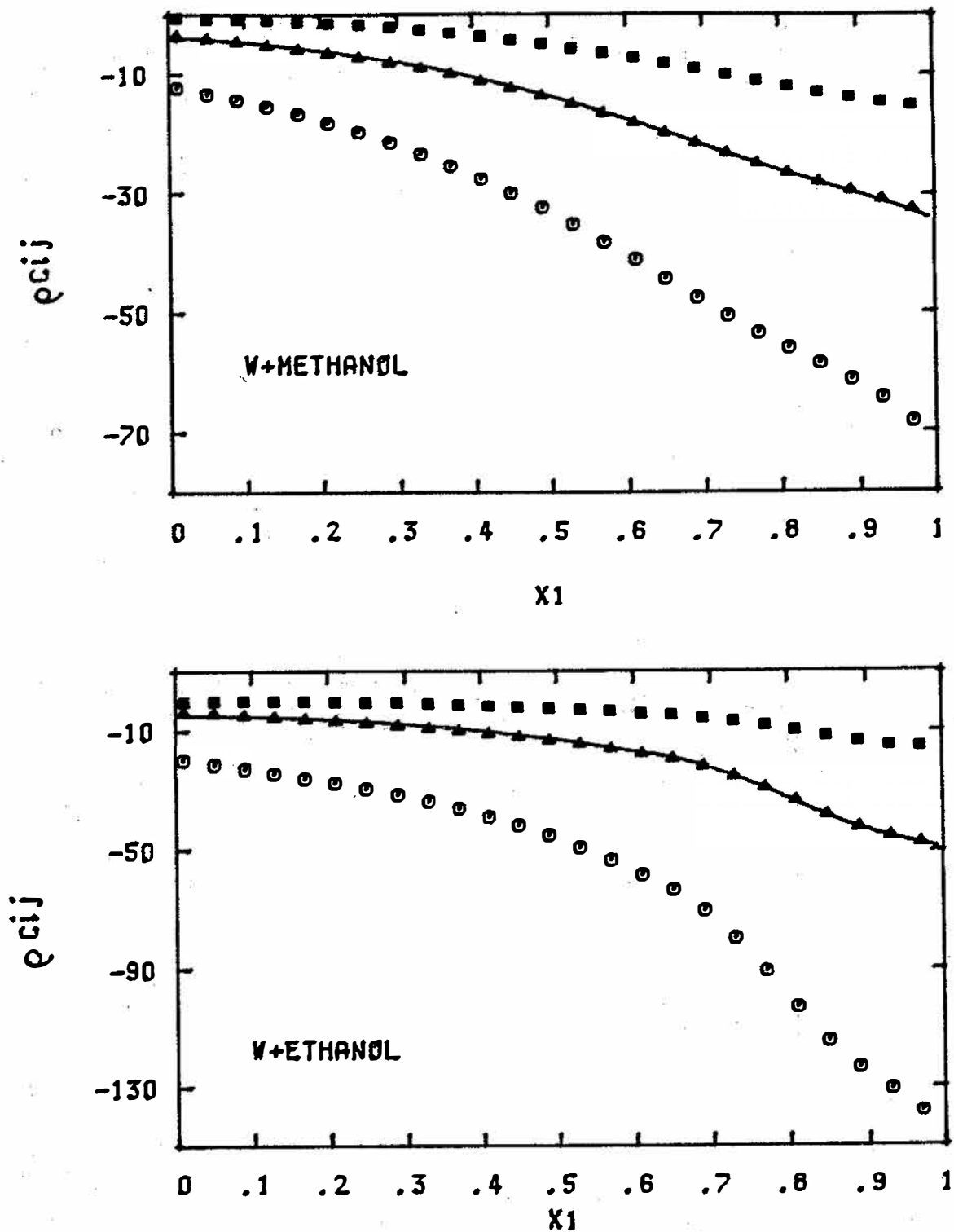

Figure 3. The Variation of $\rho C_{ij}$ with composition for water (1) + methanol (2) and for water (1) + ethanol (2) at 25 °C. The points represent experimental $\rho C_{12}$ (triangle), $\rho C_{11}$ (square), and $\rho C_{22}$ (circle) as calculated from data reported by Matteoli and Lepori (1984), Lepori and Matteoli (1988), Lepori, et al. (1988), Berti, et al. (1989) and Hamad, et al. (1989). The solid curve (———) represents the $\rho C_{12}$ prediction by the present theory using weighted arithmetic-mean $C_{12}$ closure.

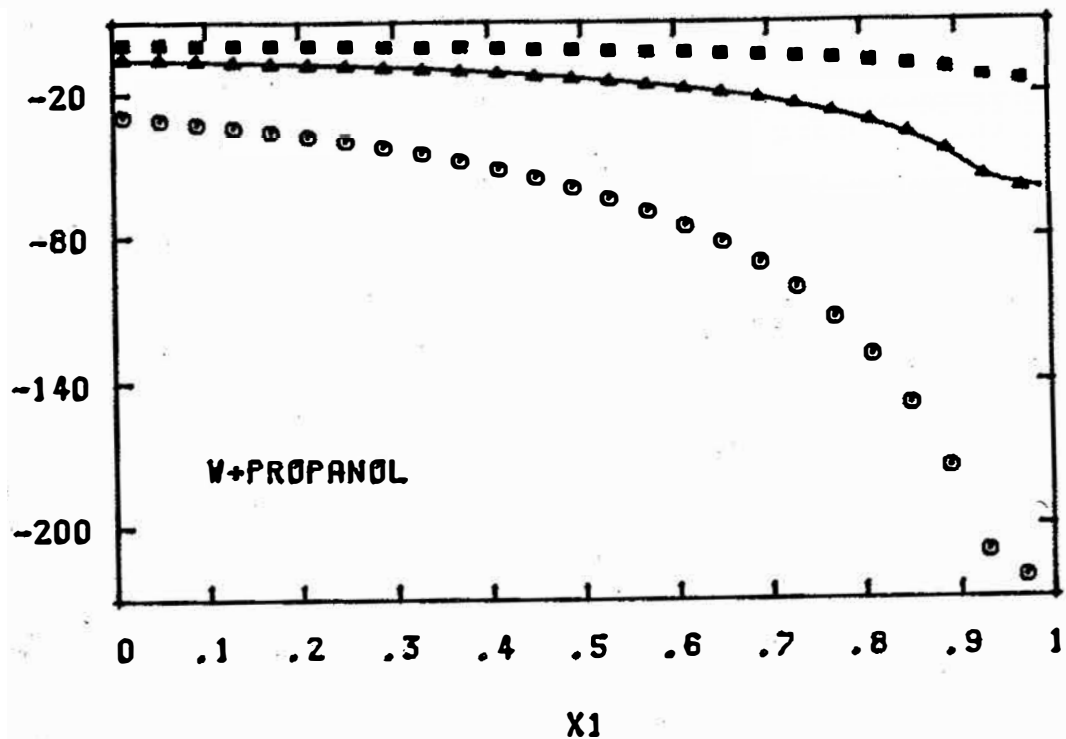

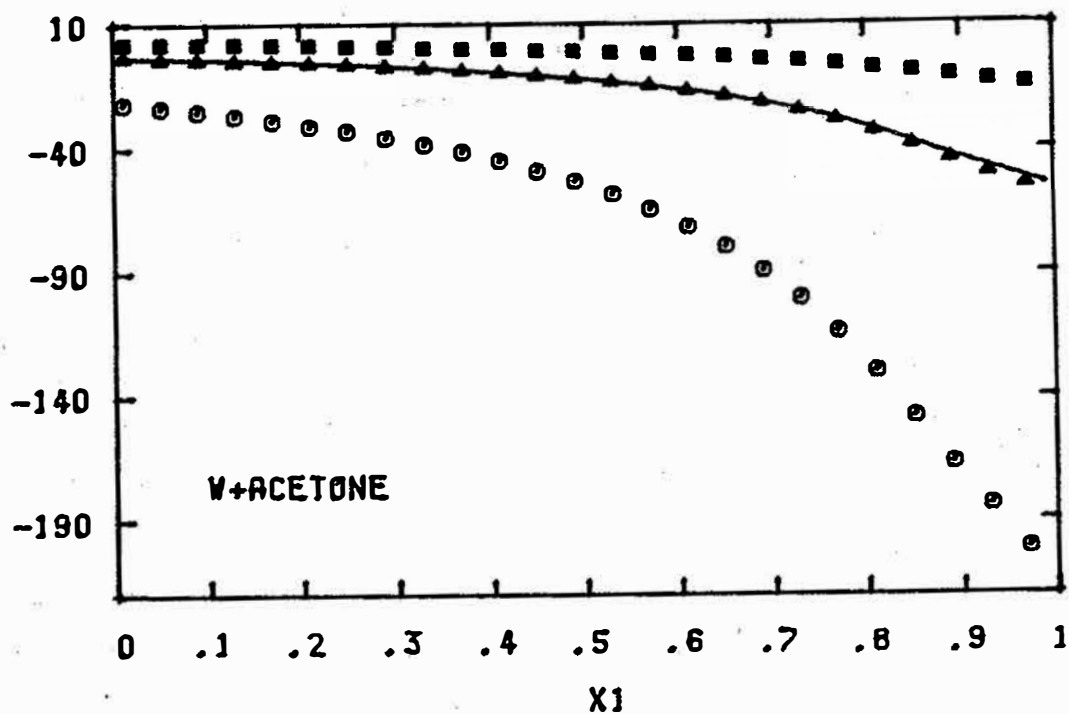

Figure 4. The variation of $\rho C_{ij}$ with composition for water (1) + propanol (2) and for water (1) + acetone (2) at 25 °C. For explanation see caption to Figure 3.

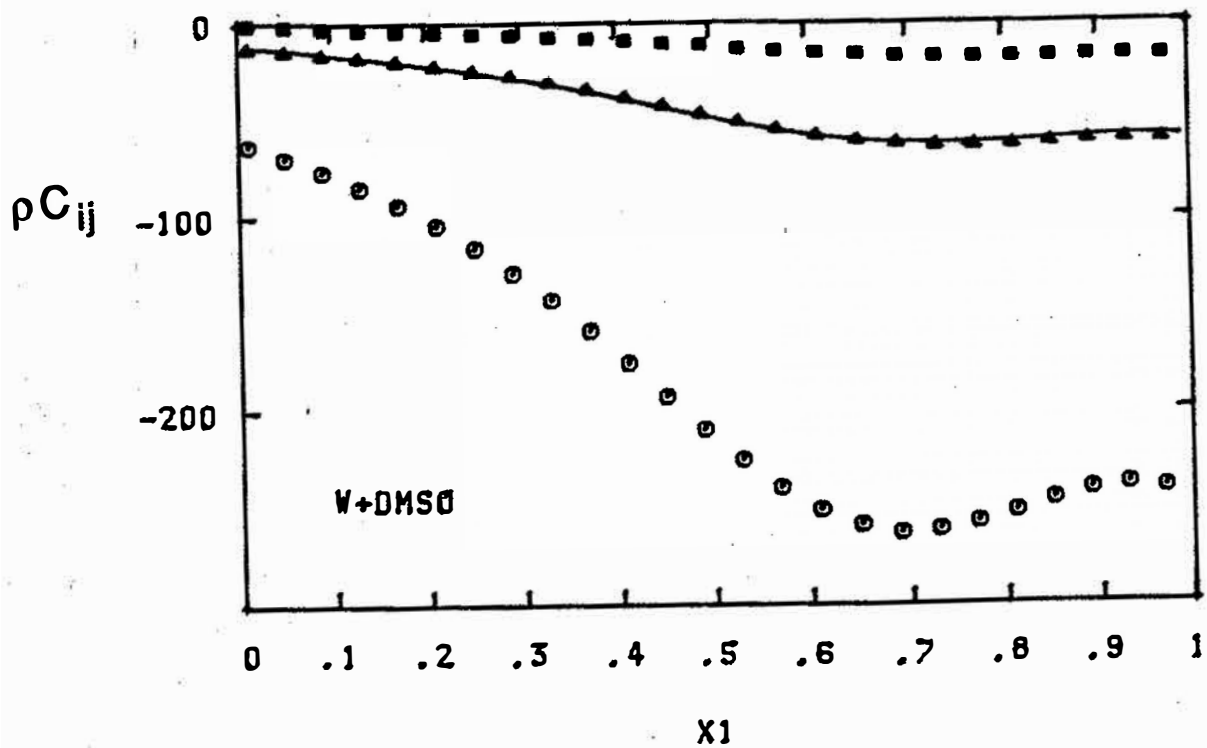

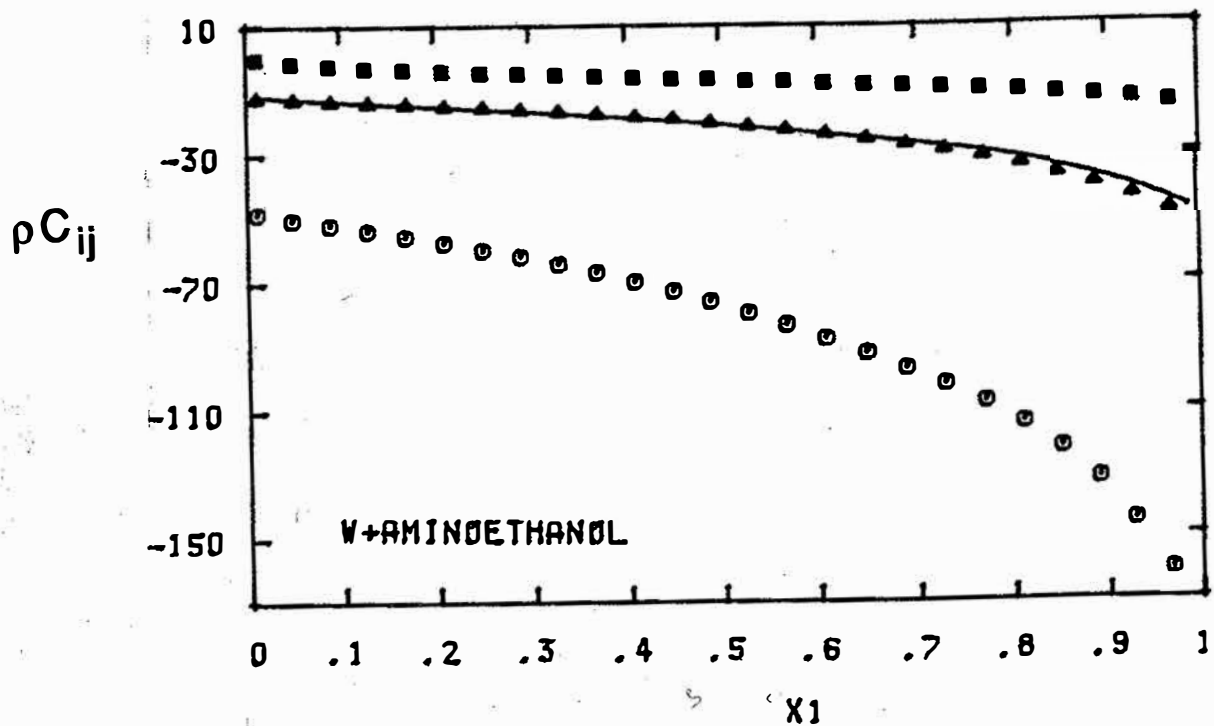

Figure 5. The Variation of $\rho C_{ij}$ with composition for water (1) + dimethylsulphoxide (DMSO) (2) and for water (1) + aminethanol (2) at 25 °C. For explanation see caption to Figure 3.

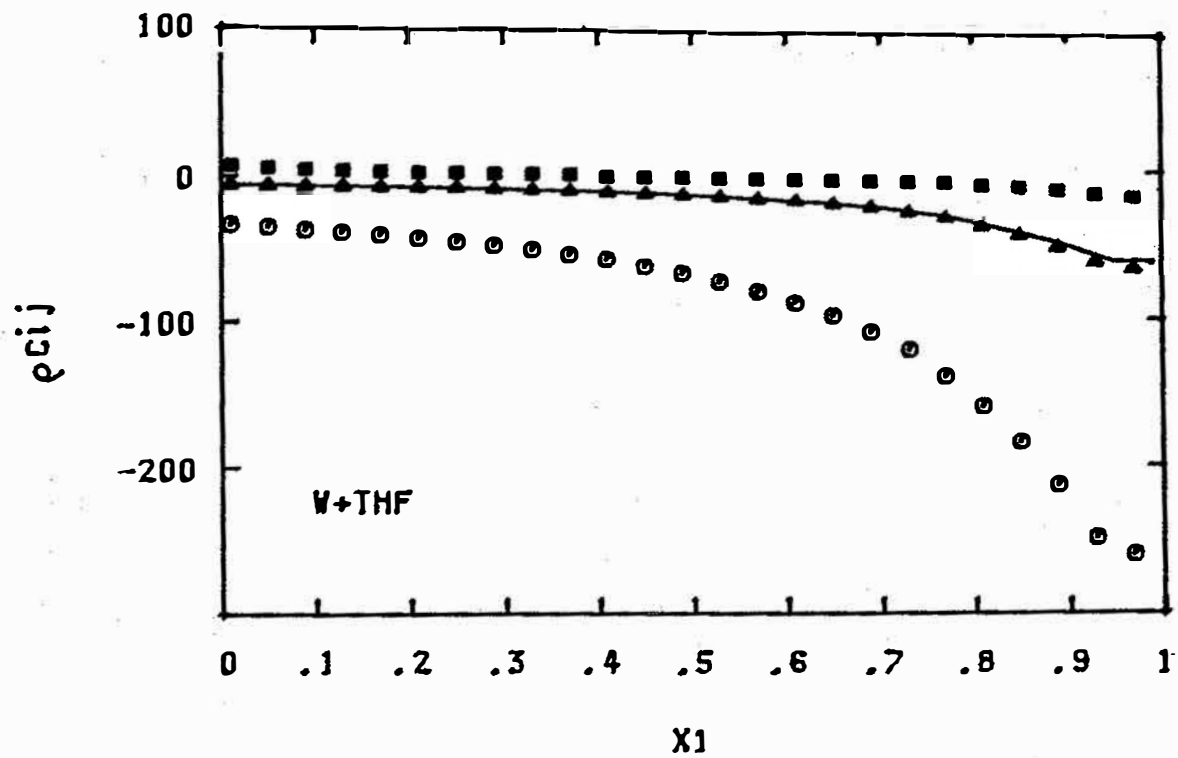

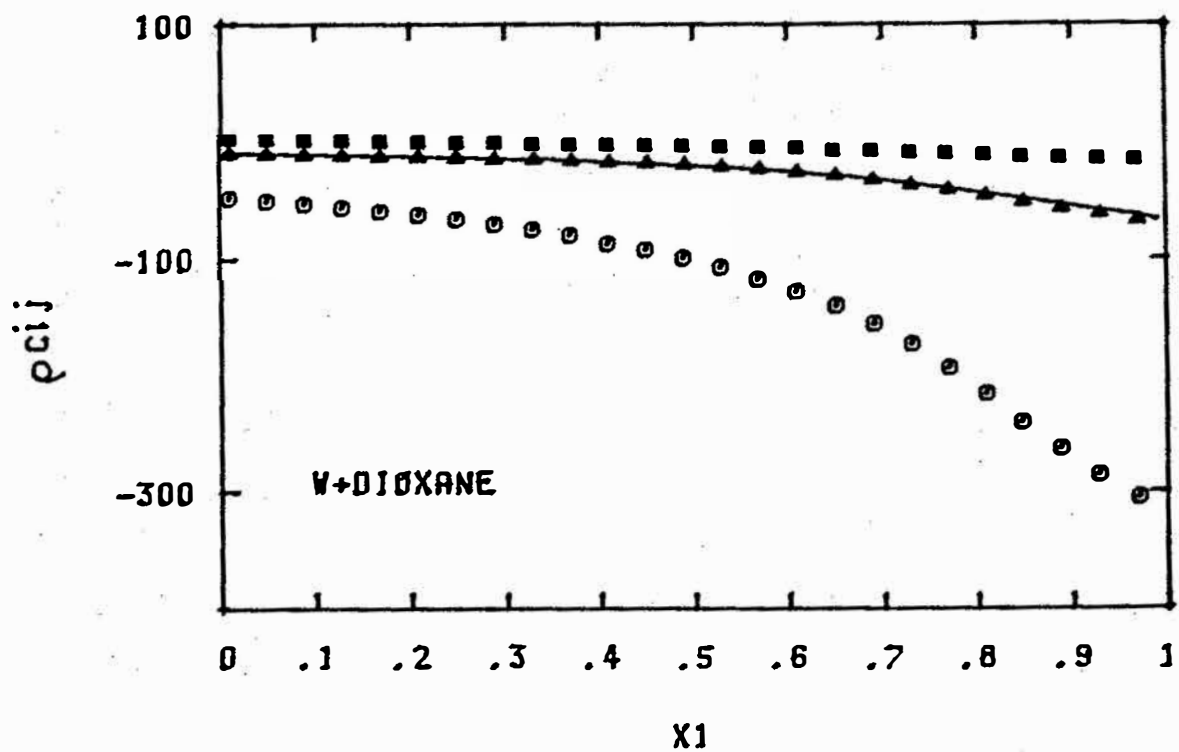

Figure 6. The Variation of $\rho C_{ij}$ with composition for water (1) + tetrahydrofuran (THF) (2) and for water (1) + dioxane (2) at 25 °C. For explanation see caption to Figure 3.

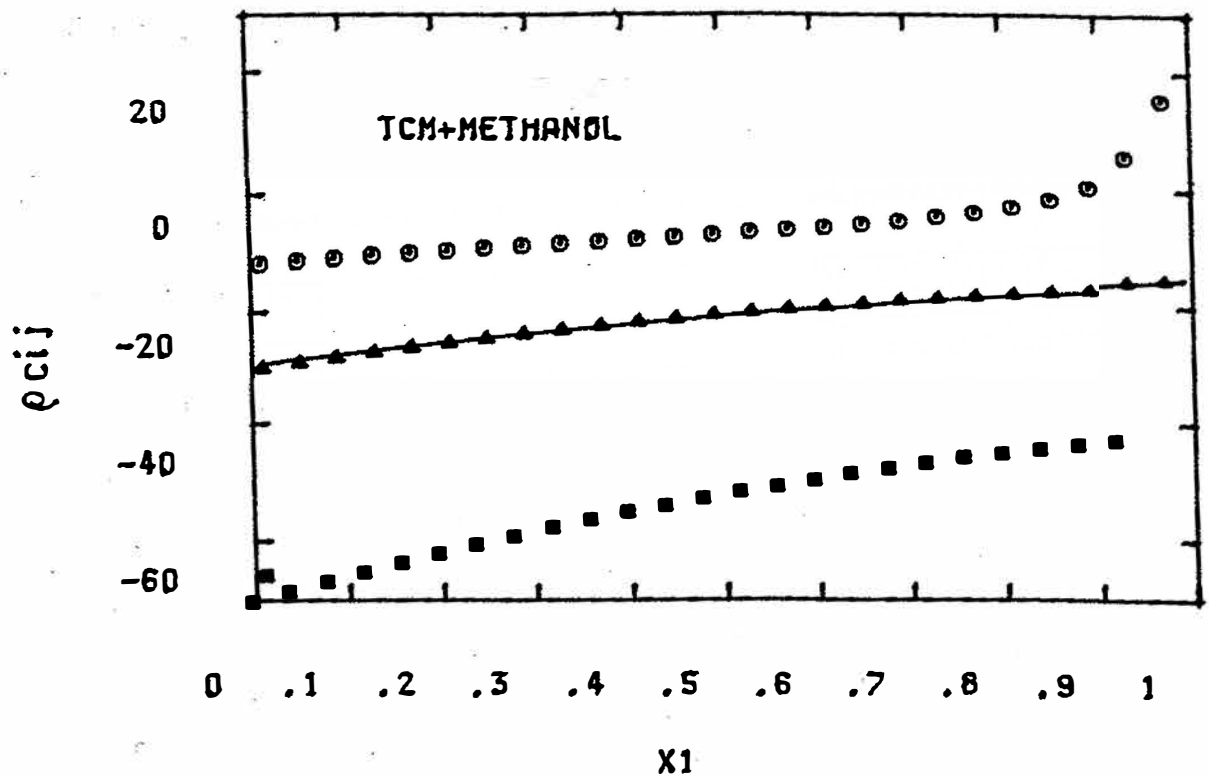

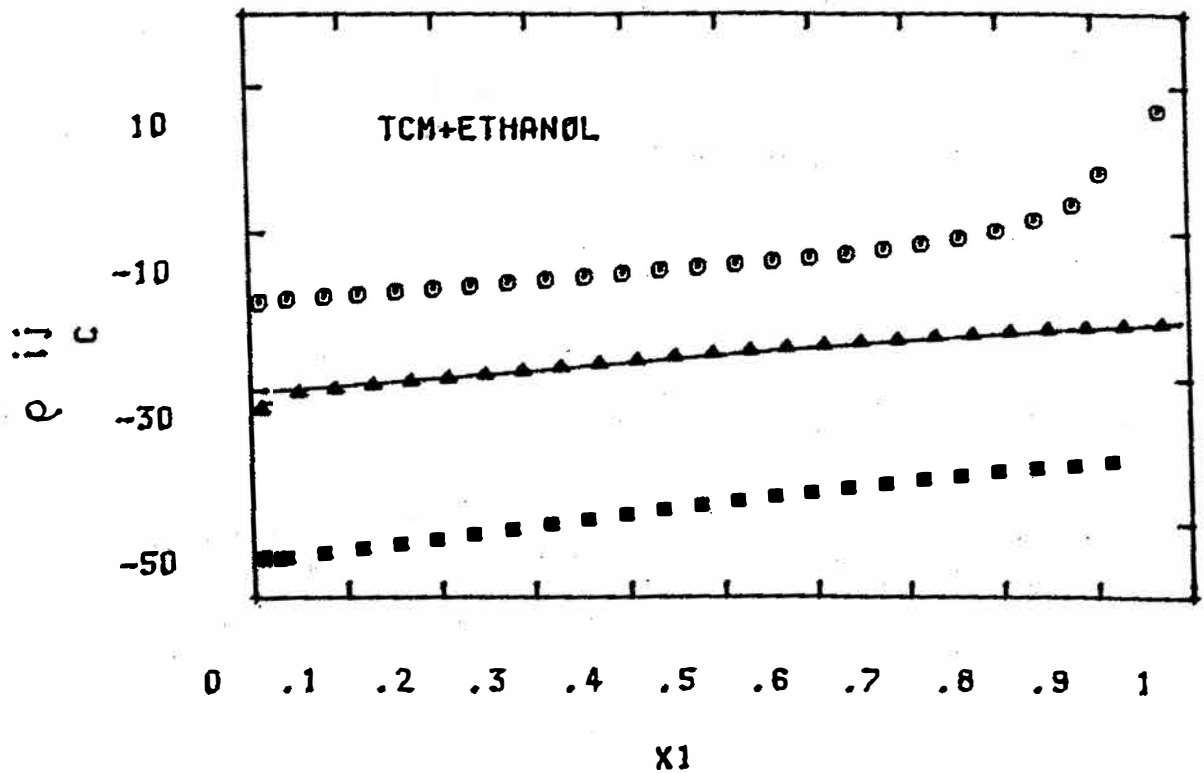

Figure 7. The Variation of $\rho C_{ij}$ with composition for tetrachloromethane (TCM) (1) + methanol (2) and for TCM (1) + ethanol (2) at 25 °C. For explanation see caption to Figure 3.

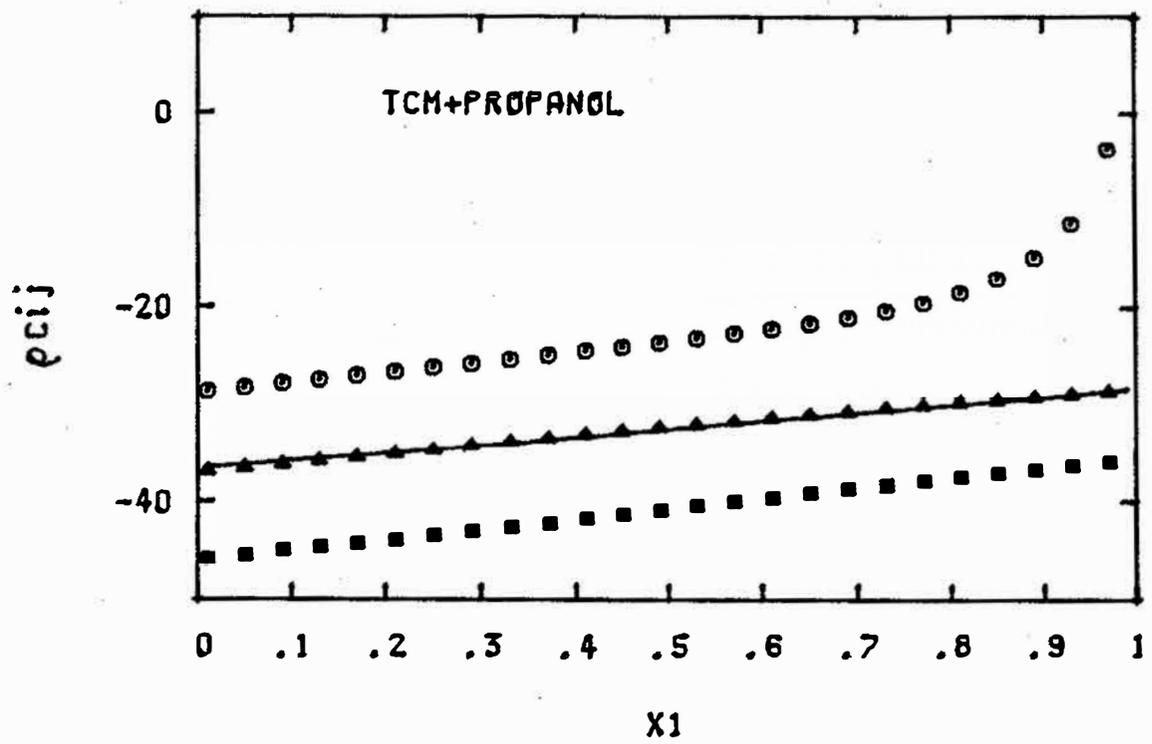

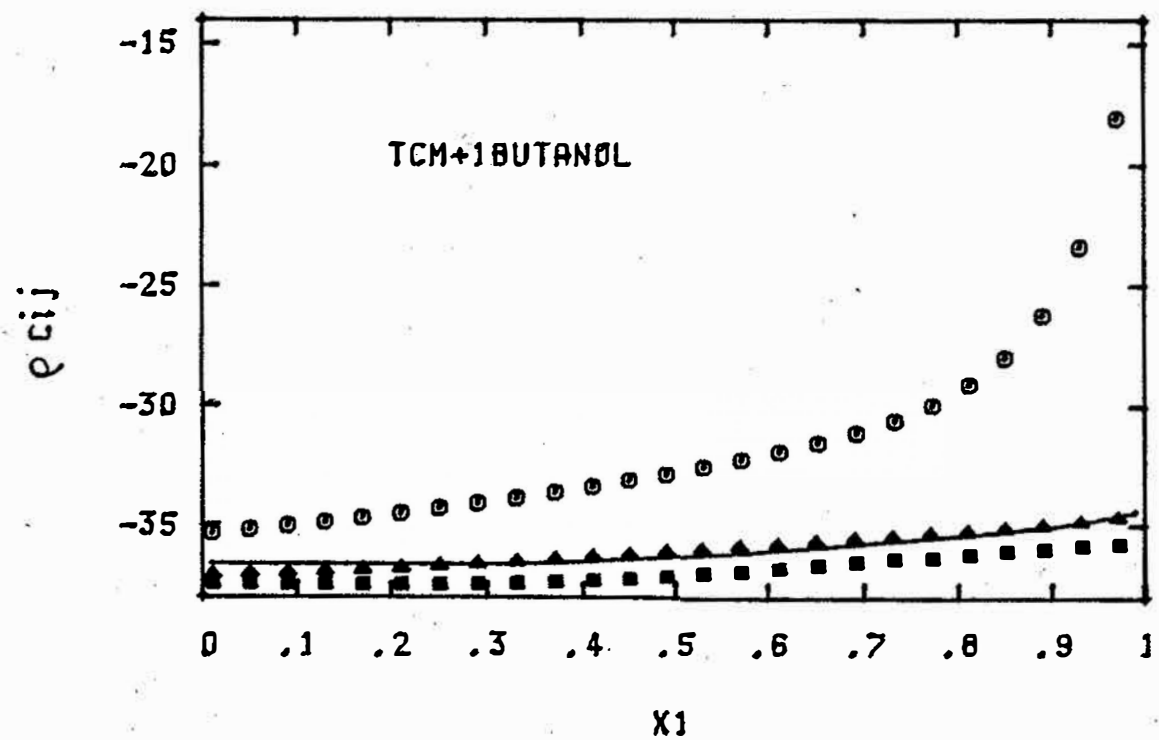

Figure 8. The Variation of $\rho C_{ij}$ with composition for TCM (1) + propanol (2) and for TCM (1) + 1-butanol (2) at 25 °C. For explanation see caption to Figure 3.

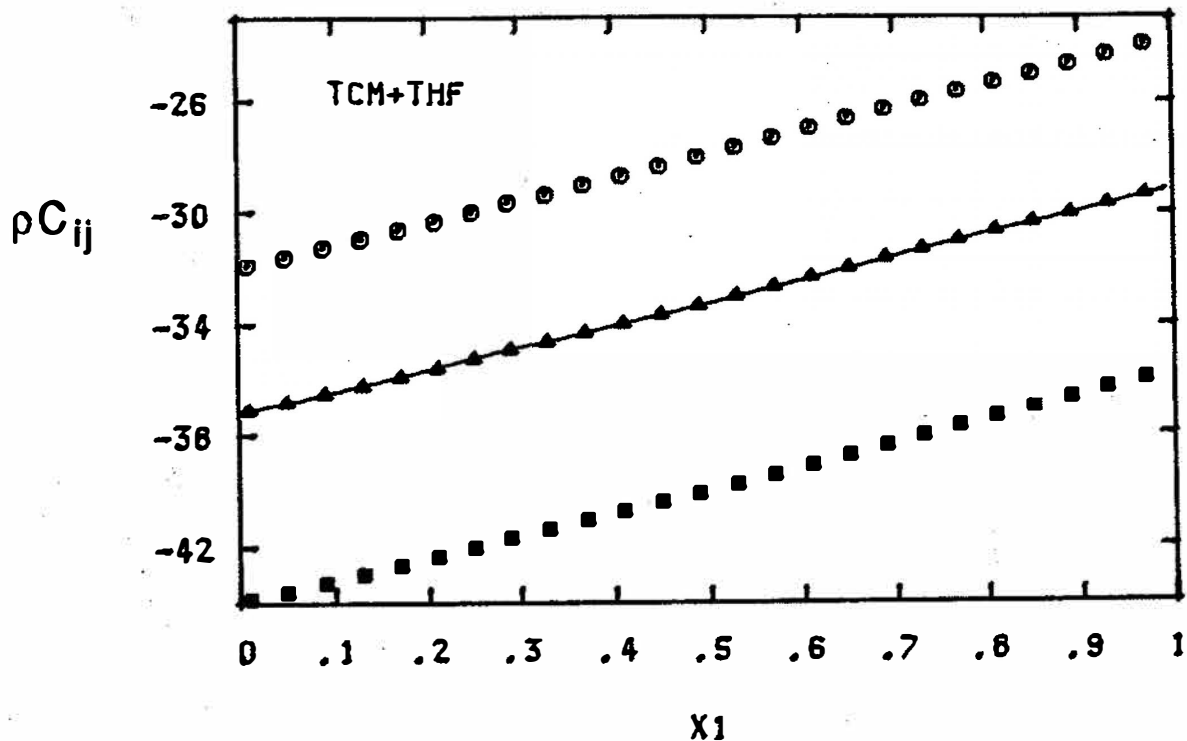

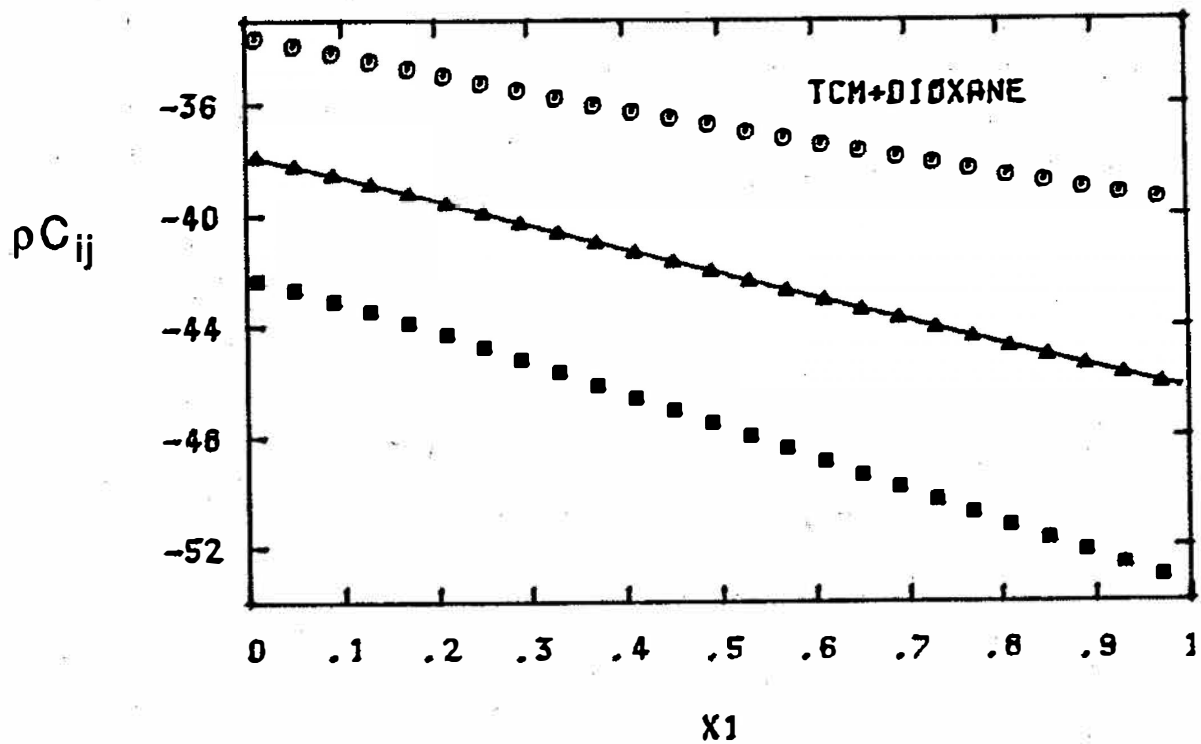

Figure 9. The Variation of $\rho C_{ij}$ with composition for TCM (1) + THF (2) and for TCM (1) + dioxane (2) at 25 °C. For explanation see caption to Figure 3.

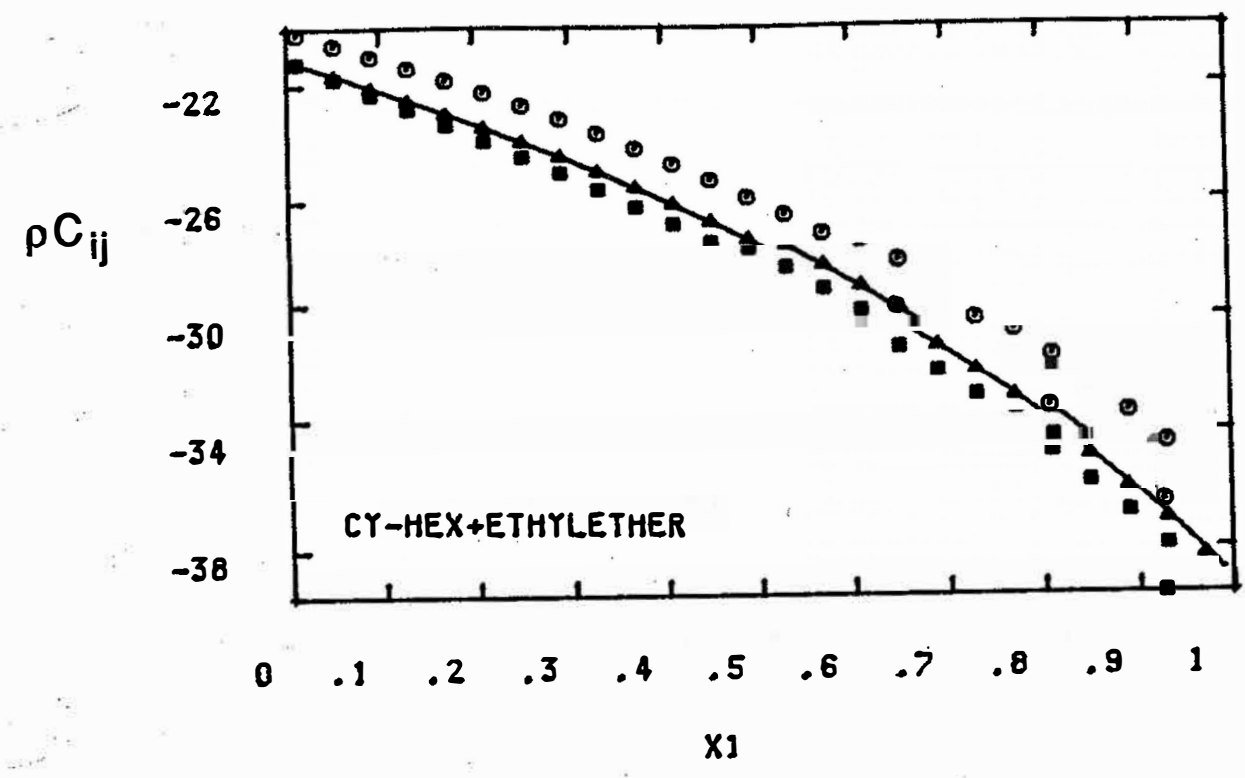
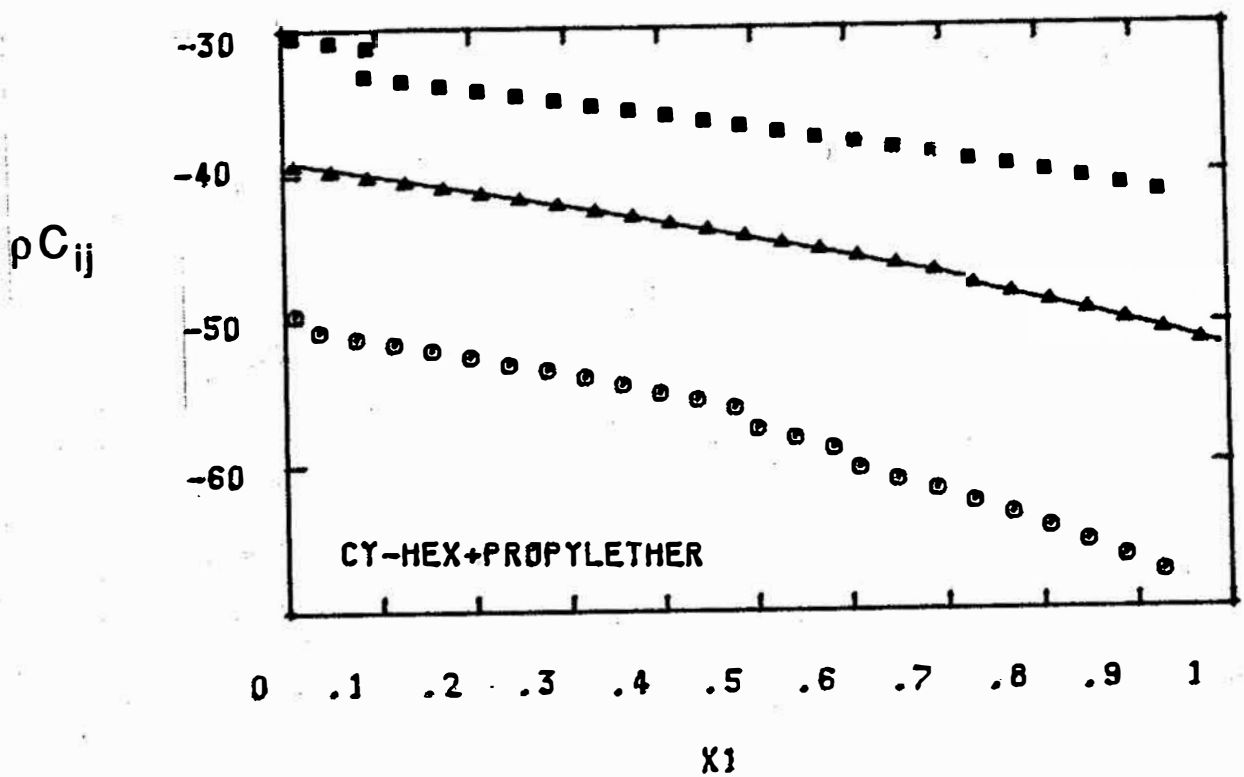

Figure 10. The Variation of $\rho C_{ij}$ with composition for cyclohexane (C-HEX) (1) + ethylether (2) and for cyclohexane (1) + propylether (2) at 25 °C. For explanation see caption to Figure 3.

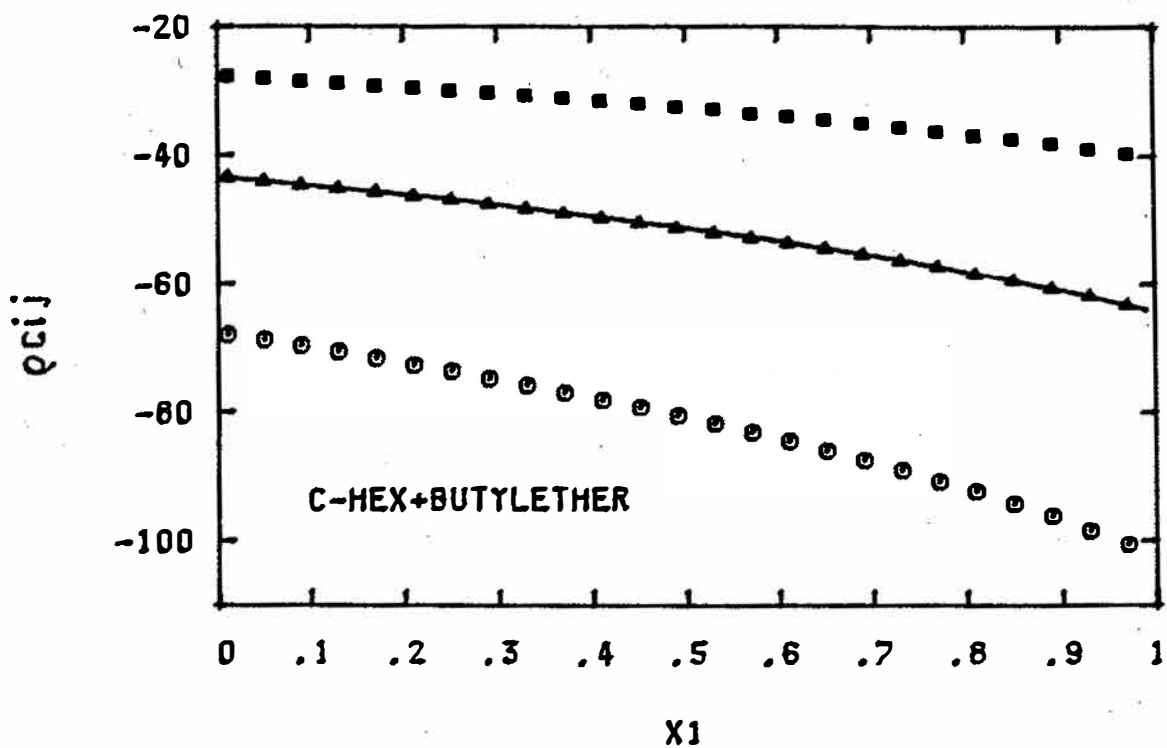

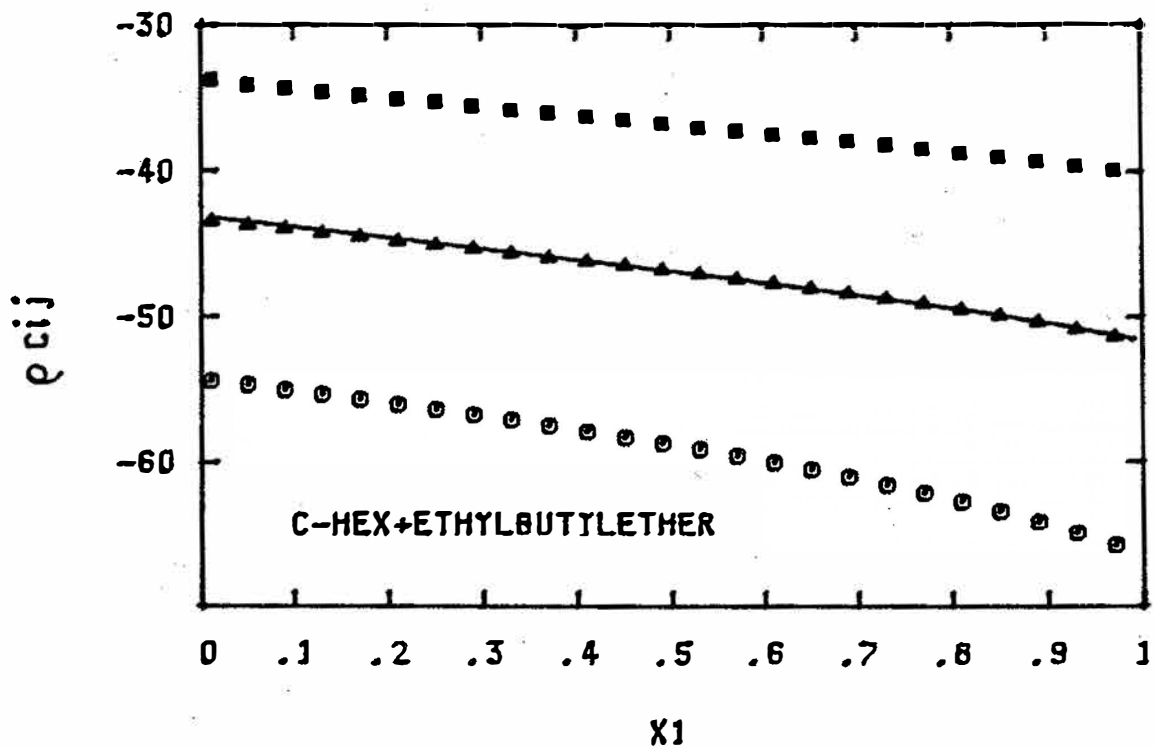

Figure 11. The Variation of $\rho C_{ij}$ with composition for cyclohexane (C-HEX) (1) + butylether (2) and for cyclohexane (1) + ethylbutylether (2) at 25 °C. For explanation see caption to Figure 3.

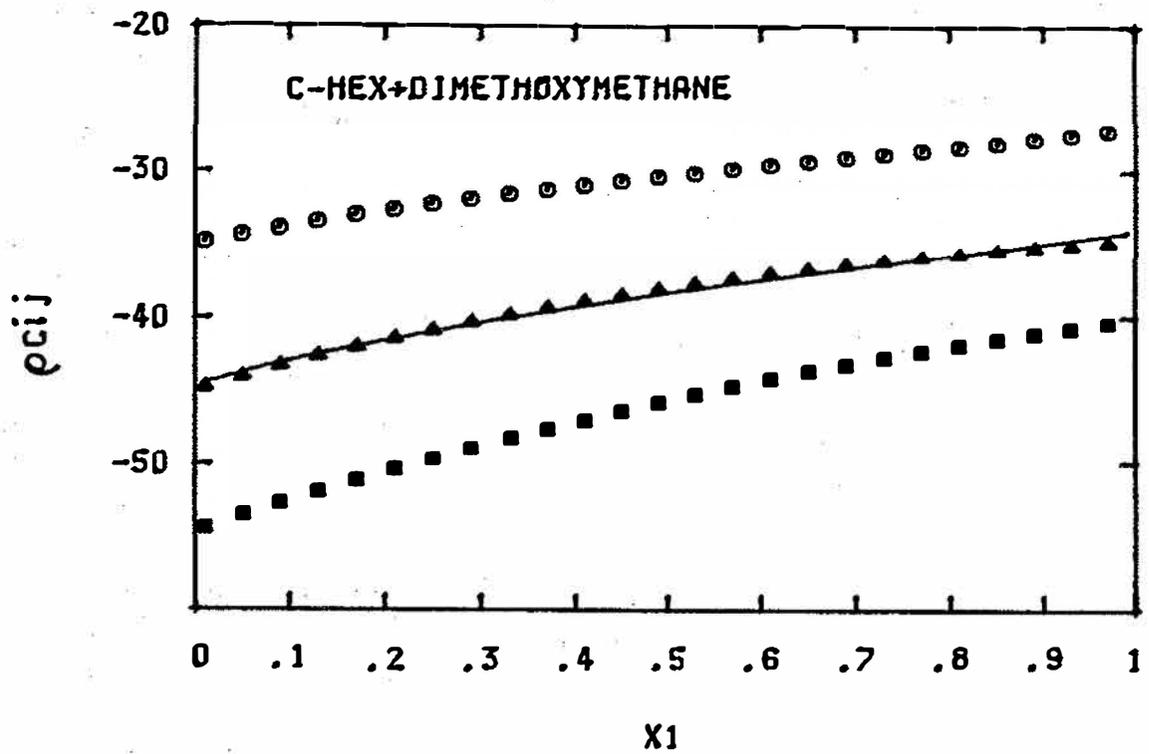

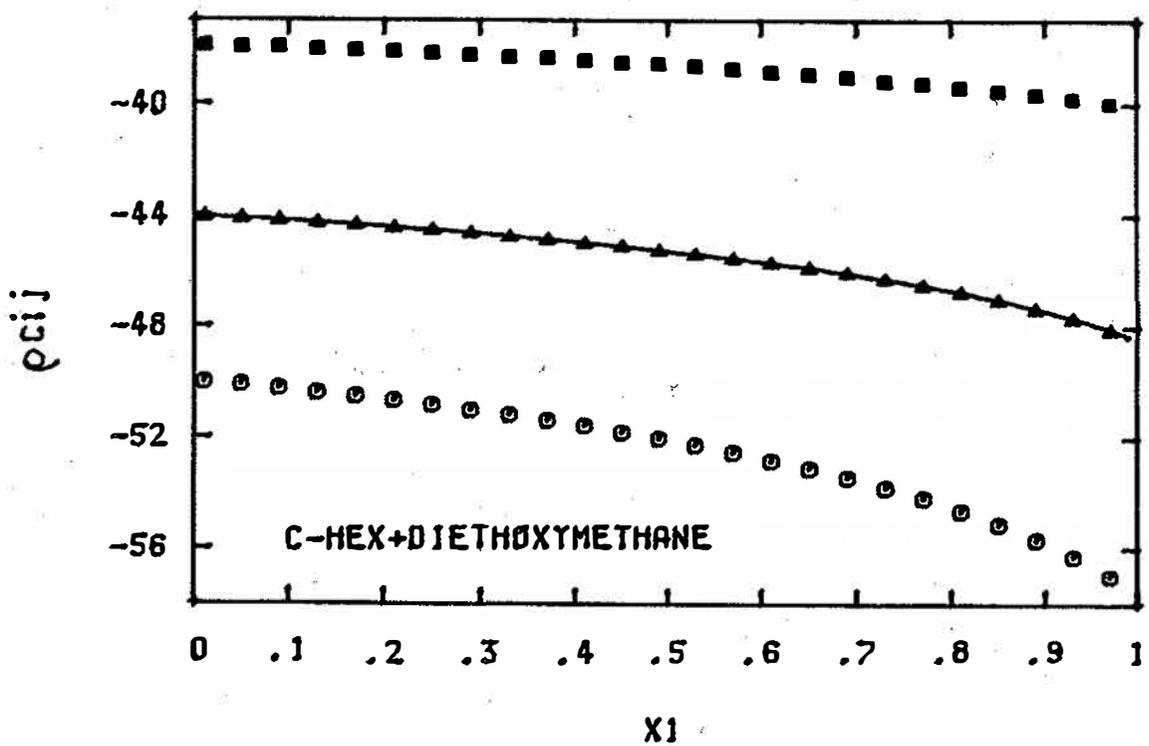

Figure 12. The Variation of $\rho C_{ij}$ with composition for cyclohexane (C-HEX) (1) + dimethoxymethane (2) and for cyclohexane (1) + diethoxymethane (2) at 25 °C. For explanation see caption to Figure 3.

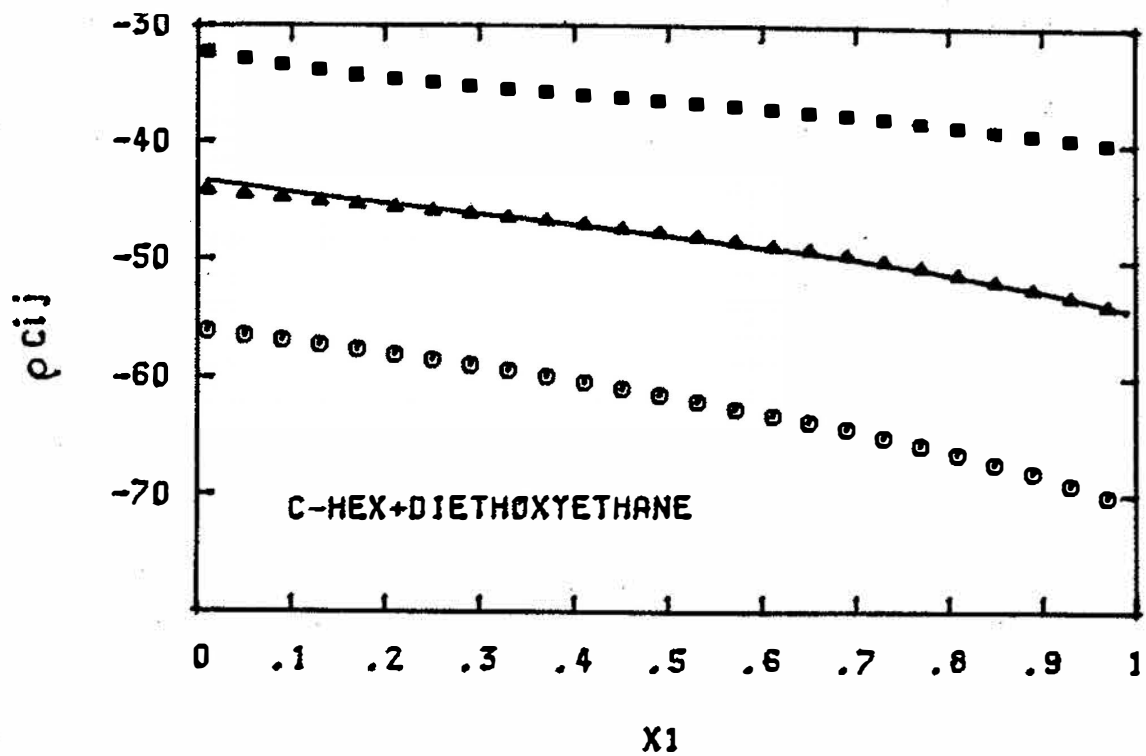

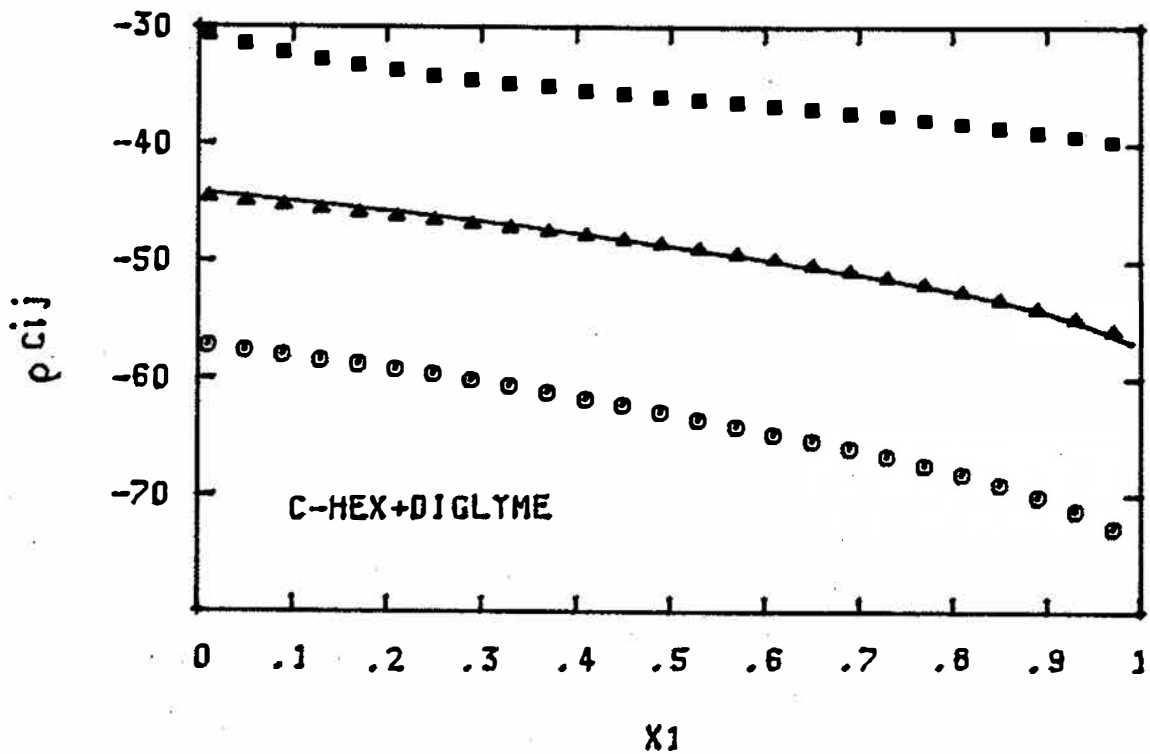

Figure 13. The Variation of $\rho C_{ij}$ with composition for cyclohexane (C-HEX) (1) + diethoxyethane (2) and for cyclohexane (1) + diethylenglycoldimethylether (diglyme) (2) at 25 °C. For explanation see caption to Figure 3.

Table 5. Molar volumes and isothermal compressibilities of the pure liquids used in this report.

| liquid | $v_i$ [cm$^3$ mol$^{-1}$] | T [°C] | $\kappa_T$ [10$^{-4}$MPa$^{-1}$] |
| --- | --- | --- | --- |
| water | 18.07 | 0.0 | 5.01 |
|  |  | 25 | 4.43 |
|  |  | 50 | 4.40 |
|  |  | 60 | 4.43 |
|  |  | 90 | 4.68 |
| tetrachloromethane | 96.50 | 25 | 10.67 |
| methanol | 40.73 | 0.0 | 10.62 |
|  |  | 25 | 12.52 |
|  |  | 60 | 15.70 |
| ethanol | 58.68 | 25 | 12.7 |
|  |  | 50 | 13.7 |
|  |  | 90 | 18.0 |
| propanol | 75.14 | 25 | 10.2 |
| n-butanol | 91.53 | 25 | 10.2 |
| tert-butanol | 94.88 | 25 | 12.0 |
| THF | 91.53 | 25 | 10.2 |
| DMSO | 70.94 | 25 | 4.60 |
| 1,4-dioxane | 85.25 | 25 | 7.20 |
| 2-aminoethanol | 60.01 | 25 | 5.0 |
| acetone | 74.05 | 25 | 12.39 |
| acetonitrile | 52.25 | 30 | 11.1 |

Data obtained as reported in Matteoli and Lepori (1984) o taken from Brostow and Maynadier (1979) or Weast et al, (1989).

For the systems water+alcohol, both $\alpha_{21}$ and $\alpha_{12}$ decrease with increasing the alcohol chain length, at the same temperature. The opposite trend is seen for the $\alpha_{21}$ and no trend for $\alpha_{12}$ of the TCM+alcohol systems. The parameter $\alpha_{21}$ and $\alpha_{12}$ show monotonic variation with temperature for water+methanol, but not for water + ethanol.

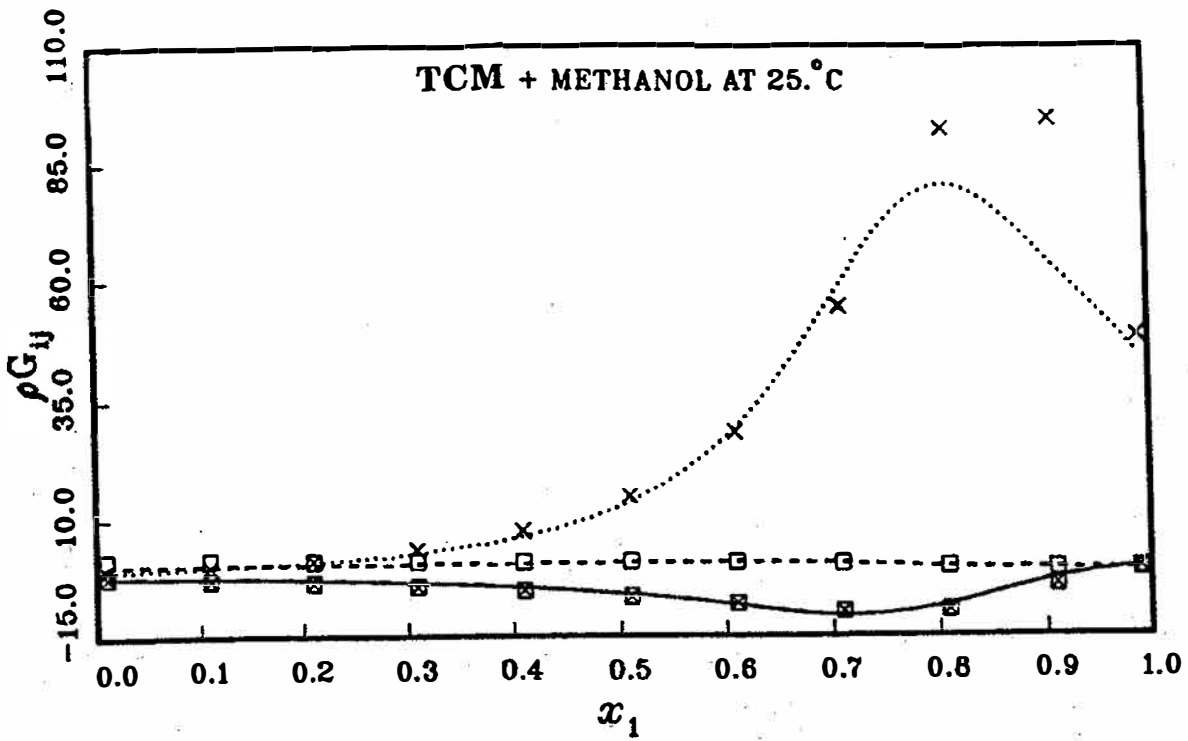

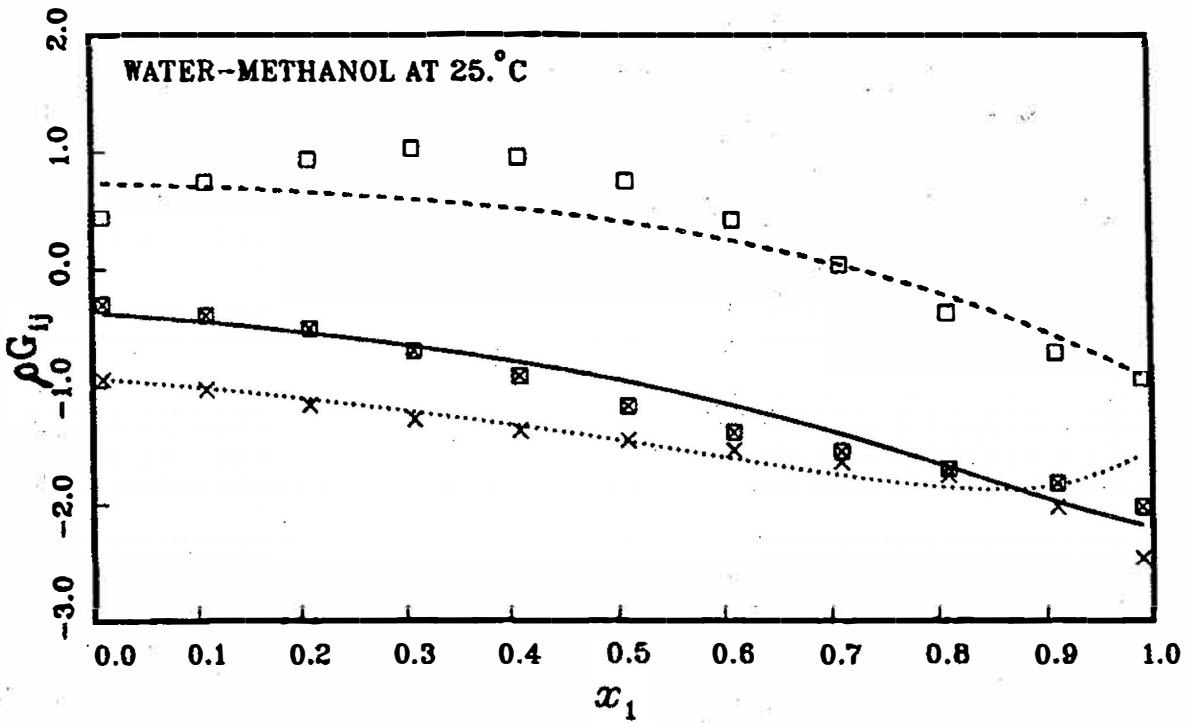

Figure 14. The Variation of $\rho G_{ij}$ with composition for tetrachloromethane (TCM) (1) + methanol (2) and for water (1) + methanol (2) at 25 °C. The points represent the experimental data (Matteoli and Lepori 1984; Lepori and Matteoli 1988; Hamad, et al. 1989): □, $\rho G_{11}$, ⊠, $\rho G_{12}$ and X, $\rho G_{22}$. The curves ( ---- $\rho G_{11}$, ——— $\rho G_{12}$, and ....... $\rho G_{22}$) represent the predictions by the present theory using weighted arithmetic-mean $C_{12}$ closure.

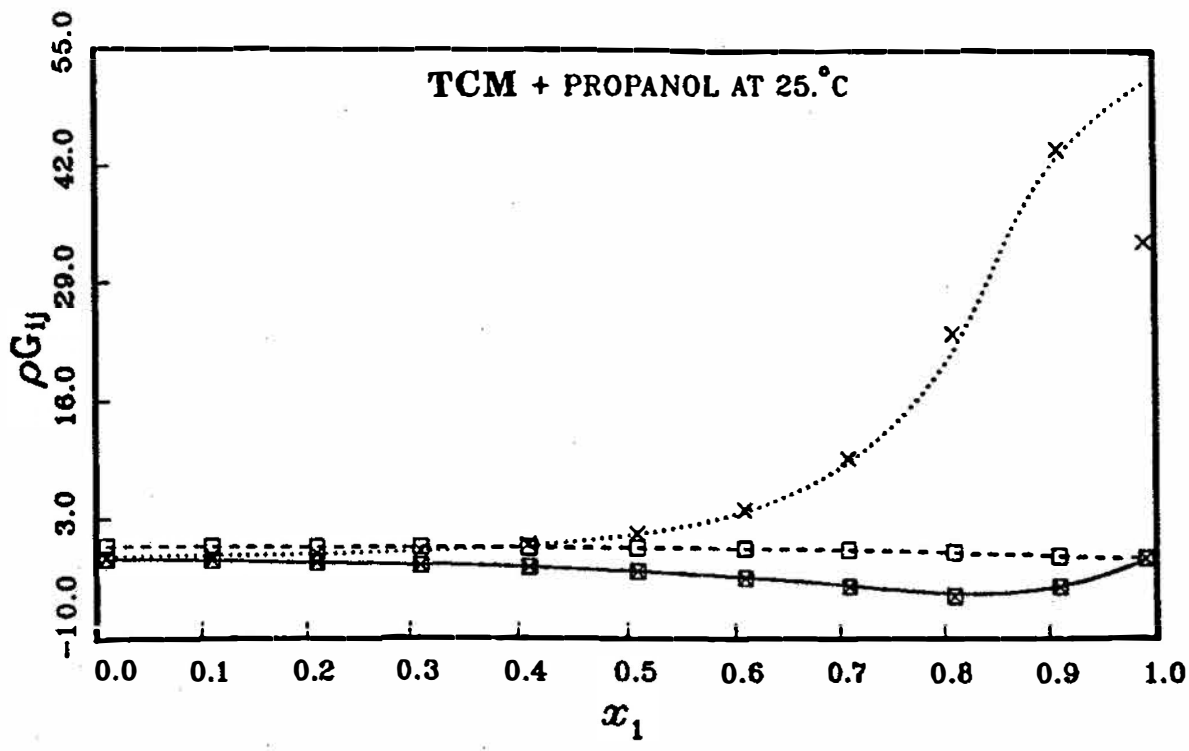

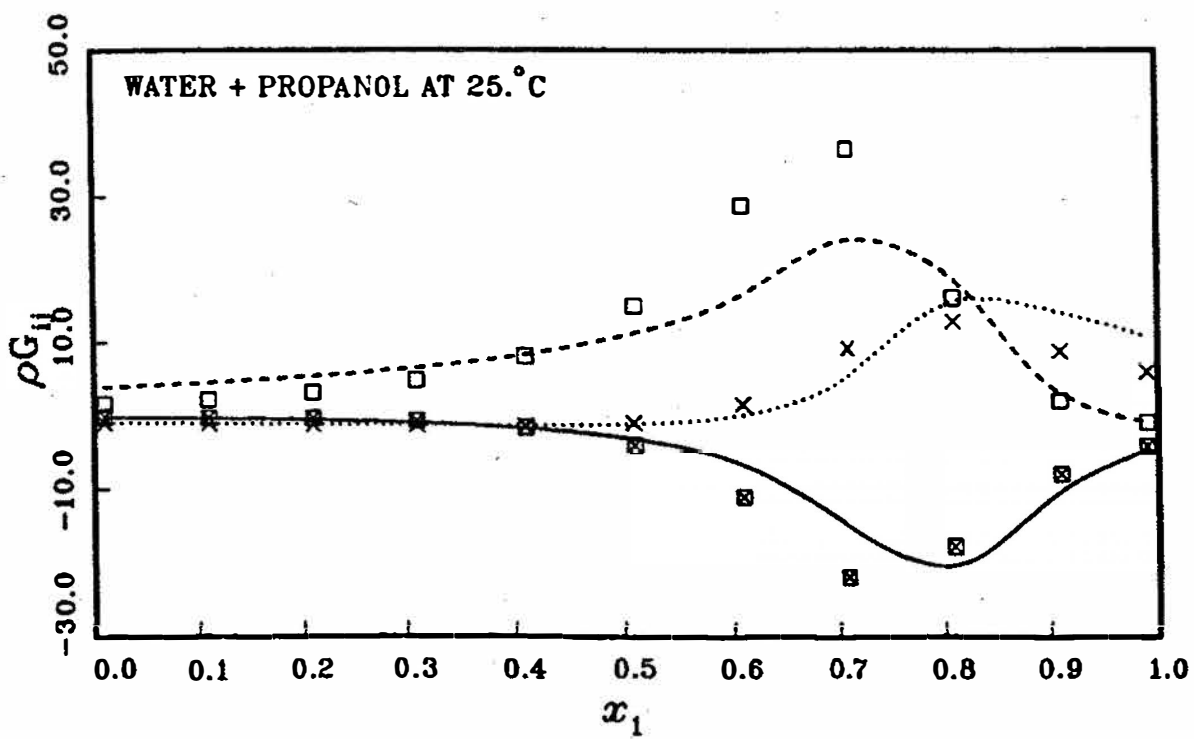

Figure 15. The Variation of $\rho G_{ij}$ with composition for tetrachloromethane (TCM) (1) + propanol (2) and for water (1) + propanol (2) at 25 °C. For explanation see caption to Figure 14.

**L. Lepori, *et al.***
Relations between and Estimations of Fluctuation Integrals and Direct Correlation Function Integrals
*Fluctuation Theory of Mixture, E. Matteoli and G.A. Mansoori (Ed's), Adv. Therm. 2, Taylor & Francis, pp.175-209, 1990*

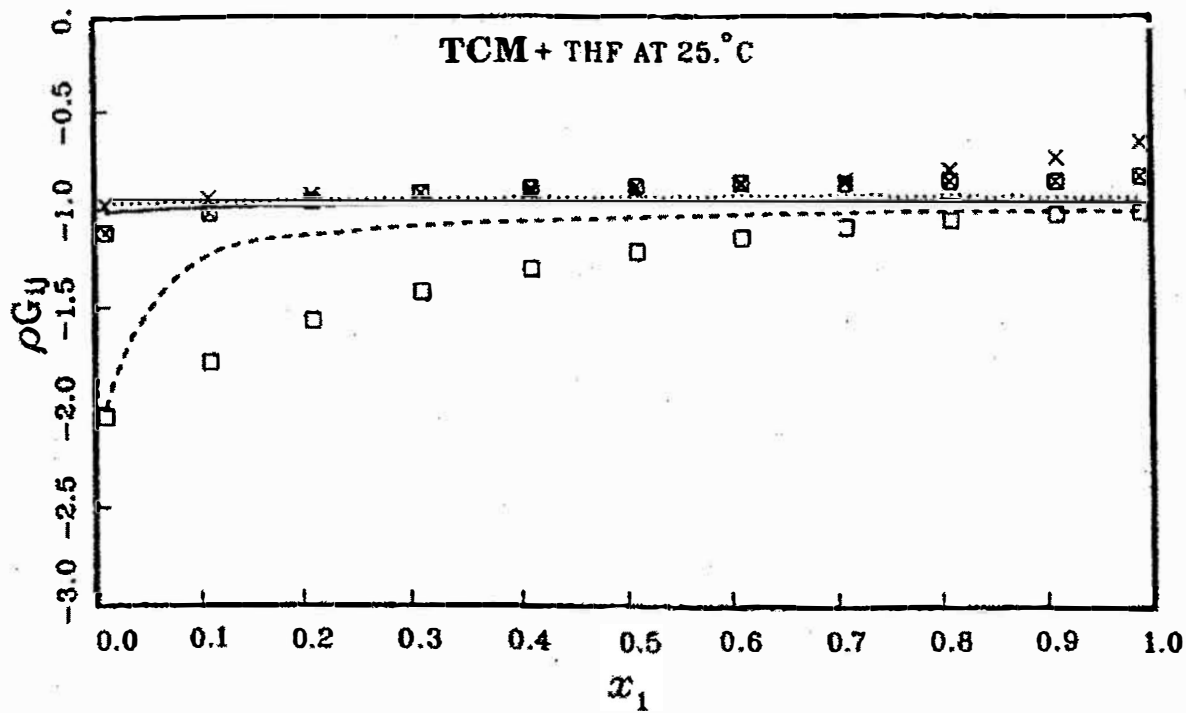

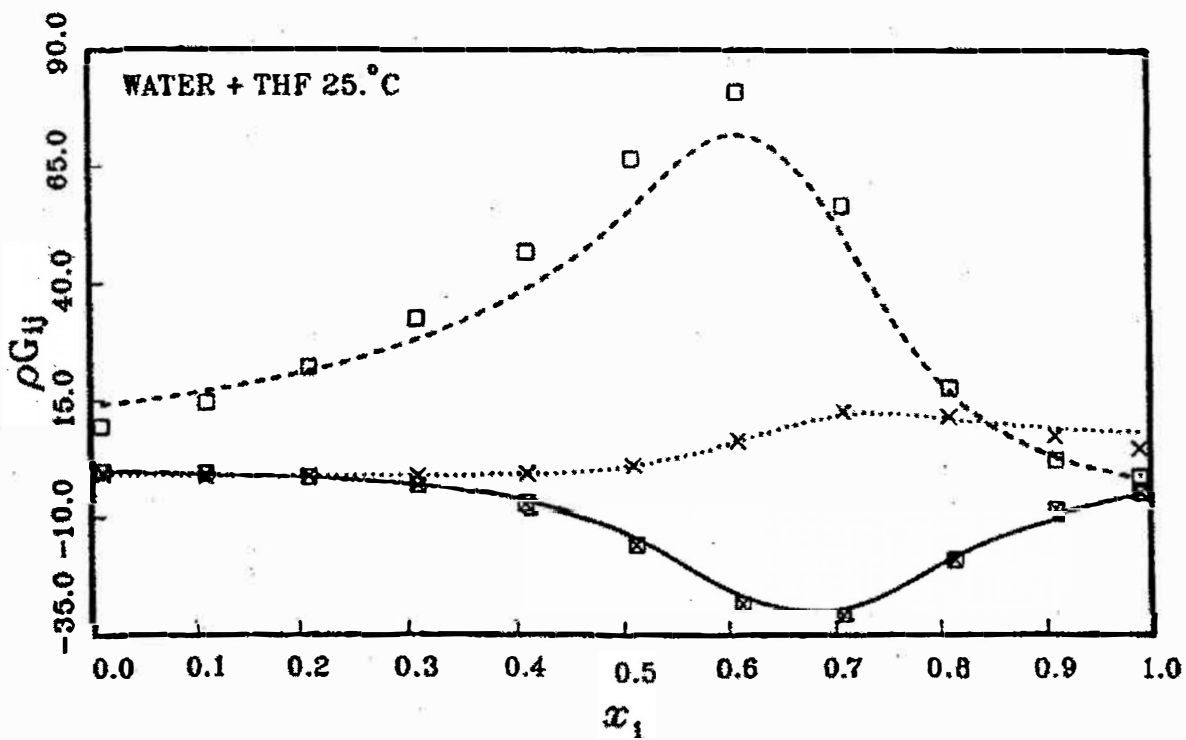

Figure 16. The Variation of $\rho G_{ij}$ with composition for tetrachloromethane (TCM) (1) + tetrahydrofuran (THF) (2) and for water (1) + THF (2) at 25 °C. For explanation see caption to Figure 14.

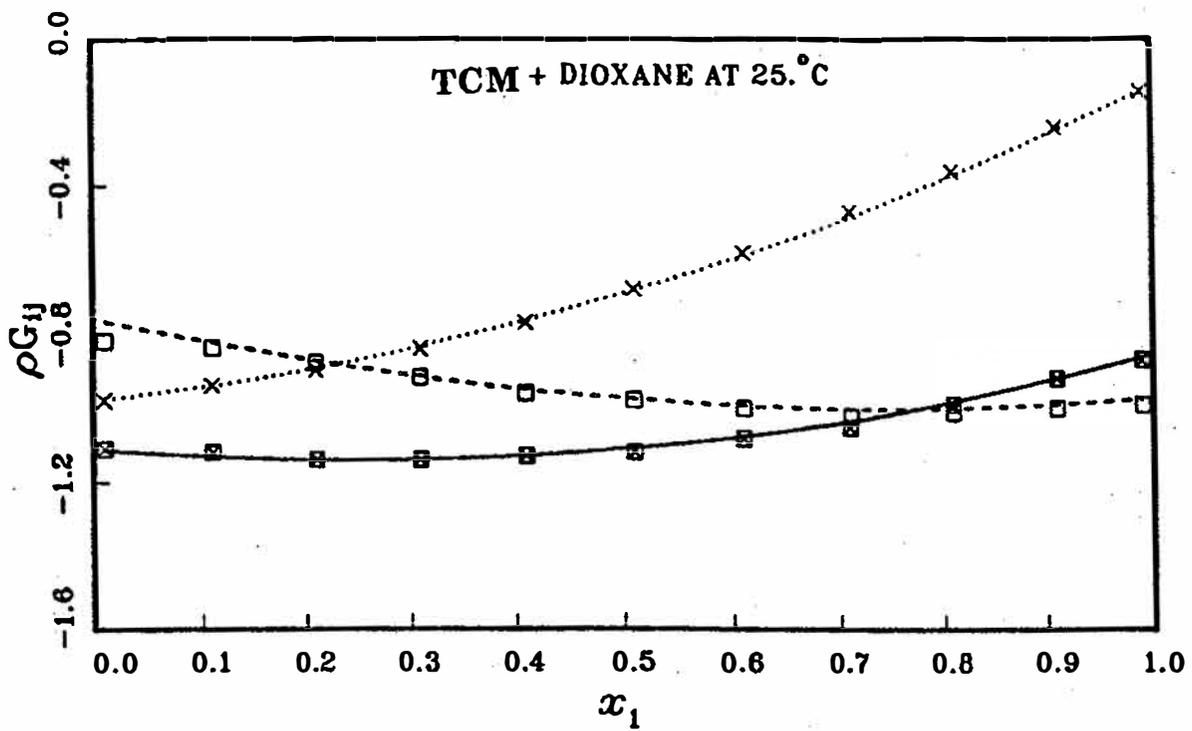

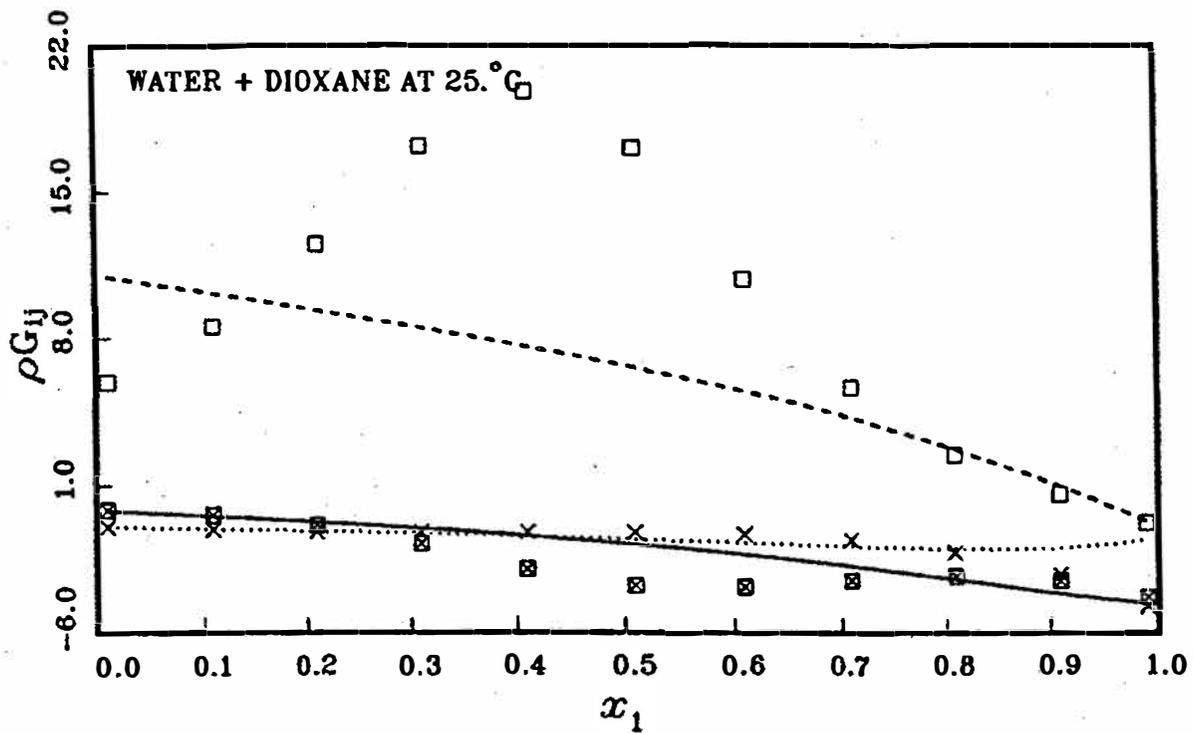

Figure 17. The Variation of $\rho G_{ij}$ with composition for tetrachloromethane (TCM) (1) + Dioxane (2) and for water (1) + Dioxane (2) at 25 °C. For explanation see caption to Figure 14.

L. Lepori, *et al.*
Relations between and Estimations of Fluctuation Integrals and Direct Correlation Function Integrals
*Fluctuation Theory of Mixture, E. Matteoli and G.A. Mansoori (Ed's), Adv. Therm. 2, Taylor & Francis, pp.175-209, 1990*

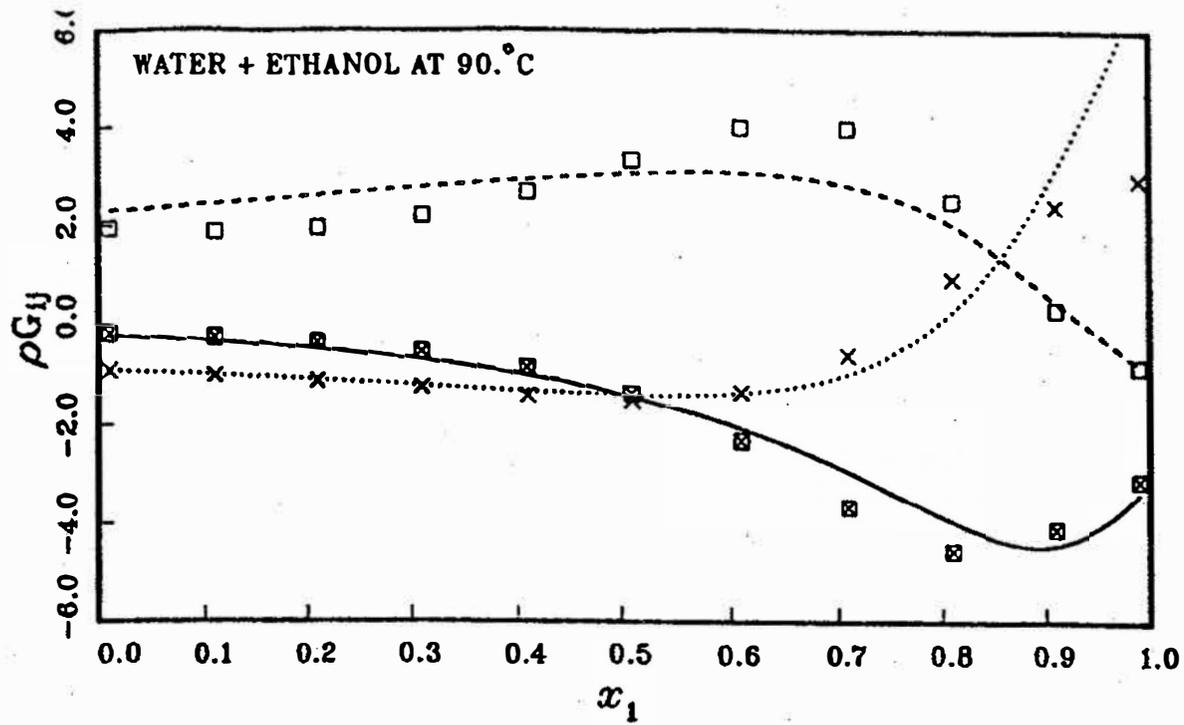

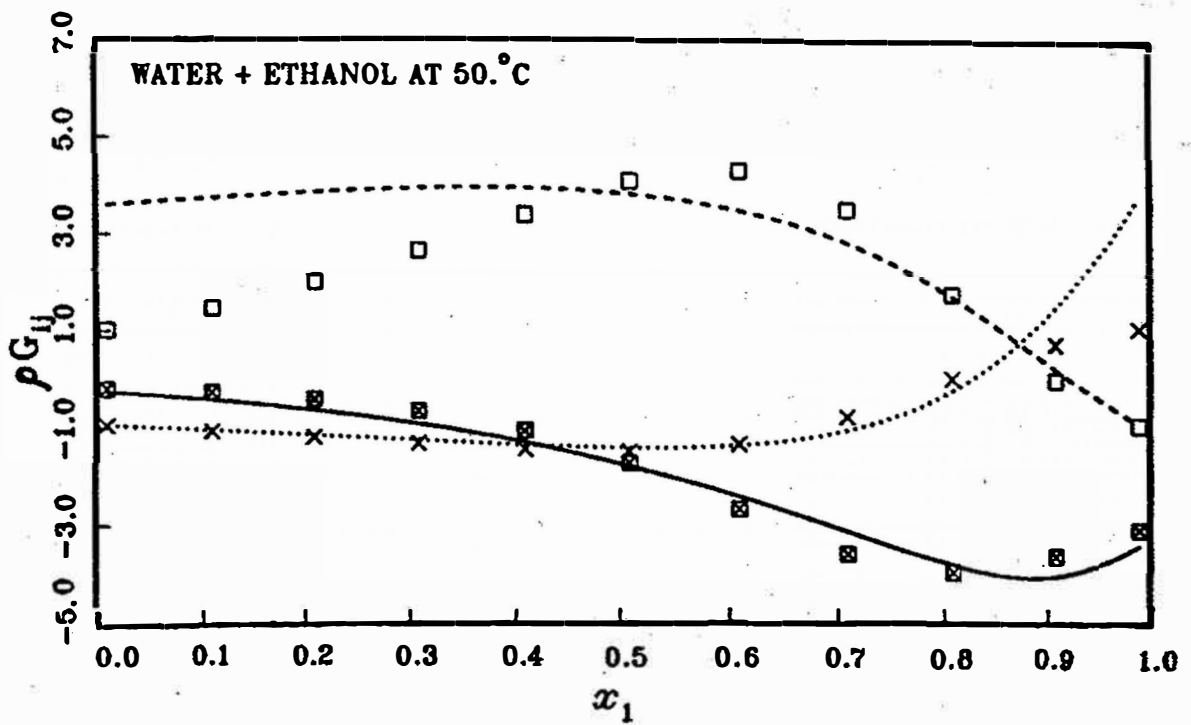

Figure 18. The variation of $\rho G_{ij}$ with composition for water (1) + ethanol (2) at two temperatures. For other details see caption to Figure 14.

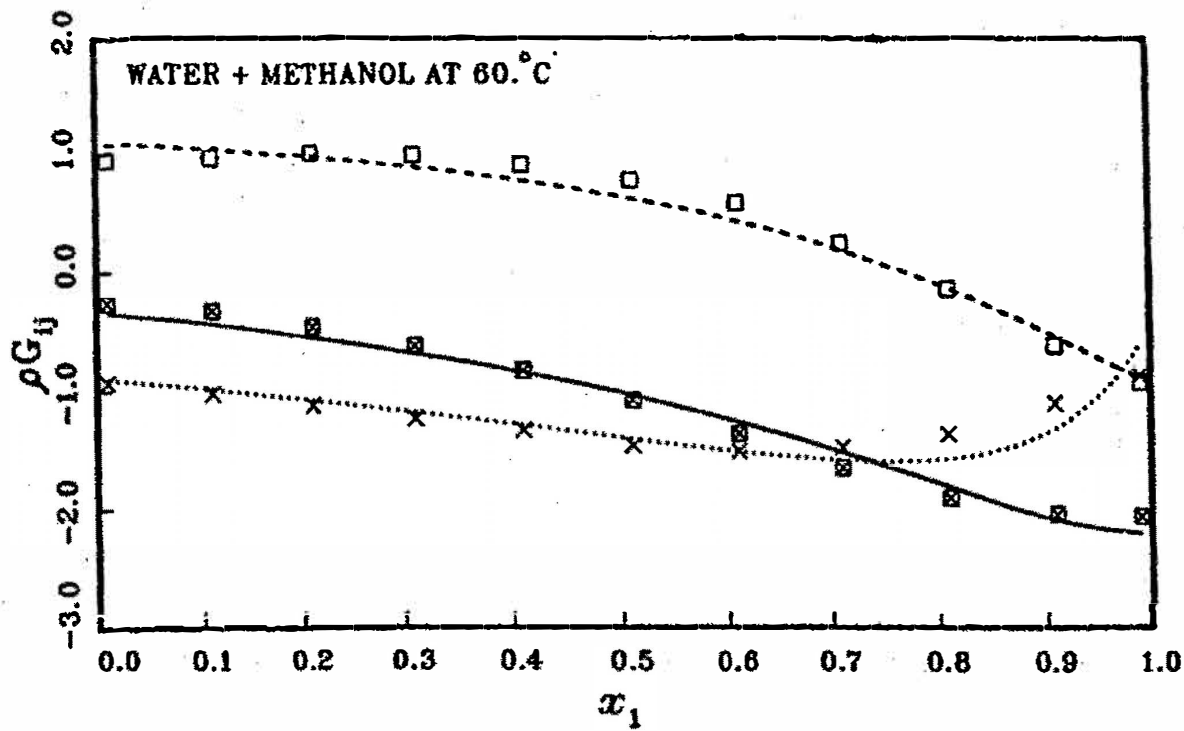

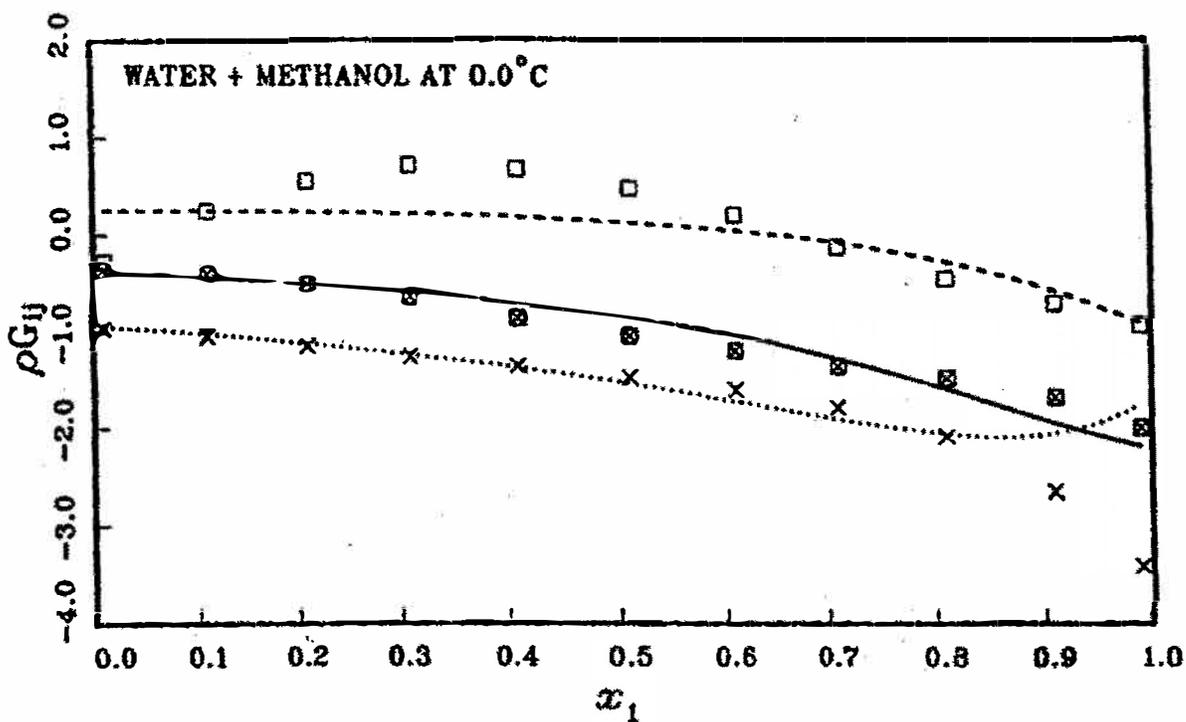

Figure 19. The variation of $\rho G_{ij}$ with composition for water (1) + methanol (2) at two temperatures. For other details see caption to Figure 14.

Table 6. Values of the interaction-parameters $\alpha_{21}$ and $\alpha_{12}$.

| organic compound | T[°C] | water + organic $\alpha_{21}$ | water + organic $\alpha_{12}$ | TCM + organic $\alpha_{21}$ | TCM + organic $\alpha_{12}$ |
|---|---|---|---|---|---|
| methanol | 0 | 0.7633 | 0.3008 | - | - |
| " | 25 | 0.4691 | 0.3577 | 0.4322 | 0.0253 |
| " | 60 | 0.3890 | 0.3790 | - | - |
| ethanol | 25 | 0.4672 | 0.2727 | 0.6161 | 0.0349 |
| " | 50 | 0.2268 | 0.3015 | - | - |
| " | 90 | 0.3494 | 0.3007 | - | - |
| propanol | 25 | 0.2657 | 0.2382 | 0.7892 | 0.0196 |
| n-butanol | 25 | - | - | 0.9317 | 0.0521 |
| tert-butanol | 25 | 0.2436 | 0.1891 | - | - |
| THF | 25 | 0.0749 | 0.2282 | 0.0182 | 1.0348 |
| DMSO | 25 | 2.4922 | 0.0932 | - | - |
| 1,4-dioxane | 25 | 0.0774 | 0.2105 | 0.6348 | 0.3241 |
| 2-aminoethanol | 25 | 1.1289 | 0.1912 | - | - |
| acetone | 25 | 0.1203 | 0.2427 | - | - |
| acetonitrile | 30 | 0.1885 | 0.3537 | - | - |

## CONCLUSION

The relations between the direct correlation function (or fluctuation) integrals reported here, Eq.s (5) & (6) when T, p, and $x_1$ are the independent variables and Eq.s (9) & (10) when T, $\rho_1$, and $\rho_2$ are the independent variables, are mathematically exact. The choice of a closure expression relating the cross direct correlation integral to the other two integrals in a binary mixture allows us to calculate these integrals simply and analytically using only the pure fluid thermodynamic data. It is demonstrated here that simple geometric and arithmetic mean closures are satisfactory for simple model fluids. It is also demonstrated that a weighted arithmetic mean closure expression is sufficient to represent the relation between the cross direct correlation integral to the other correlation integrals of binary real mixtures. Success of this theory in its application to variety of mixtures comprising of components with polar and associating intermolecular potential energies is indicative of its promise as a strong technique for prediction of properties of complex mixtures of practical interest. In the report by Hamad and Mansoori in the present monograph, applicability of this theory for prediction of phase behavior of complex mixtures is demonstrated.

ACKNOWLEDGMENT: This research is supported by the NATO Advanced Study Program.